\documentclass[aps, twocolumn, nofootinbib, longbibliography, superscriptaddress]{revtex4-2}
\usepackage{amssymb, bm, amsmath, mathtools}
\usepackage{xcolor, graphicx, url, soul, ulem}

\newcommand{\pn}{p_n}
\newcommand{\avh}{\langle h \rangle}

\newcommand{\quotes}[1]{``#1''}

\begin{document}

\title{
Alternative routes to universal diversity scaling in component systems:  from proteomes to large language models
}

\author{Andrea Mazzolini}
\email{andrea.mazzolini.90@gmail.com}
\affiliation{Department of Physics and INFN, University of Turin, via P. Giuria 1, 10125 Turin, Italy}
\affiliation{INFN, Sezione di Bari, Italy}

\author{Leonardo Agasso}
\affiliation{Department of Physics and INFN, University of Turin, via P. Giuria 1, 10125 Turin, Italy}

\author{Filippo Valle}
\affiliation{Department of Physics and INFN, University of Turin, via P. Giuria 1, 10125 Turin, Italy}

\author{Michele Caselle}
\affiliation{Department of Physics and INFN, University of Turin, via P. Giuria 1, 10125 Turin, Italy}

\author{Marco Cosentino Lagomarsino}
\affiliation{Department of Physics and INFN, University of Milan, via Celoria 16, 20139 Milan, Italy}
\affiliation{IFOM-ETS–The AIRC Institute of Molecular Oncology, The Associazione Italiana di Ricerca sul Cancro (AIRC) Institute of Molecular Oncology, Milan 20139, Italy}

\author{Matteo Osella}
\email{matteo.osella@unito.it}
\affiliation{Department of Physics and INFN, University of Turin, via P. Giuria 1, 10125 Turin, Italy}

\begin{abstract}
Remarkably common statistical laws characterize the diversity scaling and its fluctuations across a wide range of complex "component systems".  These regularities are often interpreted as signatures of an underlying innovation mechanism driving the growth of component diversity, but the basic ingredients necessary for their emergence  remain poorly understood. 
In particular, from language and technological artifacts to genomes and gene expression patterns, the number of distinct components grows sublinearly with system size, while its variance scales approximately as the square of its mean.
This behavior is consistent across diverse systems, raising the question of whether general constraints or emergent principles underlying diversity and innovation define the architectures of realizations with different numbers of components.  %
To address this question, we derive analytical conditions for the joint emergence of these two diversity laws  within a broad class of growth models, showing that they require a specific asymptotic dependence of the innovation probability on  diversity and system size. We then demonstrate that the same macroscopic laws arise in a different class of models with latent heterogeneity, where quadratic fluctuation scaling always emerges asymptotically as a consequence of  general statistical principles, essentially the law of total variance, without explicitly assuming an innovation mechanism or any specific rule for system assembly. We compare these predictions with empirical data from language, genomes, LEGO constructions, and texts generated by large language models. Our results show that empirical diversity scaling laws strongly constrain generative models but do not uniquely identify the mechanisms generating diversity, revealing a close correspondence between innovation-driven growth models and latent-variable descriptions.

\end{abstract}

\maketitle

\section{Introduction}

Diversity is a defining property of complex ``component systems"~\cite{page2010diversity,mazzolini2026componentsystems}. From genomes and cellular transcriptomes to  ecosystems, human organizations, languages and technological artifacts such as LEGO toys or software, these modular  systems are characterized by the variety of components they contain and their combinations~\cite{van2003scaling,grilli2020macroecological,lazzardi2023emergent,altmann2016statistical,koch2007software}. Component diversity underlies functionality, adaptability, resilience, and evolution, raising key questions concerning its emergence and dynamics across disciplines~\cite{quigley1998urban,whittaker1972evolution, hunter1998value, yang2026scaling,holehouse2025generativemodelfunctiongrowth}.

A common approach to this question focuses on average behavior. Across many systems, the number of distinct components (the vocabulary) increases sublinearly with system size (total number of components), a pattern known as Heaps' law in linguistics~\cite{herdan1960type, heaps1978information, altmann2016statistical, cosentino2009universal, locey2016scaling}. This regularity has motivated a large body of work aimed at identifying the mechanisms responsible for the emergence of the observed dependence of diversity on system size. 
A common interpretation is that these patterns reflect the underlying process of innovation~\cite{erwin2015novelty, hochberg2017innovation, fagerberg2006innovation}. In this view, new components are progressively discovered as the system grows, and the observed scaling laws encode the dynamics of novelty generation. This perspective motivates a large class of growth models, from Yule–Simon processes~\cite{yule1925ii} to several more recent direct generalizations~\cite{zanette2005dynamics, gerlach2013stochastic, holehouse2025generativemodelfunctiongrowth}, or analogous models based on the Chinese restaurant process~\cite{pitman1997two, cosentino2009universal}, or on the Pólya urn formalism~\cite{tria2014dynamics}. Another class of growth models assumes the presence of an underlying structure of the component space  that constrains the innovation dynamics~\cite{corominas2015understanding,mazzolini2018heaps,mazzolini2018zipf,iacopini2018network}.

An alternative framework  explains the emergence of the average scaling of Heaps' law as a  \quotes{null}  statistical consequence of the power-law component-frequency distribution  \cite{van2005formal, lu2010zipf, eliazar2011growth}, i.e. Zipf's law \cite{kingsley1935psycho}.
In this statistical description, realizations are  the result of  random sampling of components with a given probability distribution~\cite{mazzolini2026componentsystems}, and Heaps' law is simply a consequence of the underlying heterogeneous probabilities, without any explicit innovation mechanism.  

Yet, diversity is not characterized by its average alone. Realizations of comparable size can display substantially different levels of diversity, and these fluctuations themselves exhibit striking statistical regularities. 
More specifically, this "diversity of diversity" can  be characterized by looking at the variance of the vocabulary as a function of its average value. In linguistics~\cite{gerlach2014scaling} and in some  social systems~\cite{tria2020taylor} a quadratic fluctuation scaling has been  observed, a non-self-averaging scaling  reminiscent of Taylor's law in ecology \cite{taylor1961aggregation, eisler2008fluctuation, giometto2015sample,grilli2020macroecological}.
 We will show that this quadratic fluctuation  scaling  extends to very different component systems such as genomes and LEGO toys. 
Also in this case,  both  specific models based on growth with innovation~\cite{gerlach2013stochastic} and a framework based on latent variables~\cite{tria2020taylor} have been invoked to explain the fluctuation scaling.  

These observations raise a fundamental question: what gives rise to the variability of diversity across realizations in such a variety of component systems? More generally, what are the minimal ingredients required to explain both Heaps' law and its fluctuation scaling, and can these statistical laws distinguish between alternative generative mechanisms?

Here, we address these questions systematically through a mathematical modeling approach, comparing two broad explanatory frameworks: sequential growth through innovation and statistical sampling with latent heterogeneity. We first derive the conditions under which a general class of innovation models jointly reproduces Heaps' law and quadratic fluctuation scaling. We show that this requires a specific asymptotic dependence of the innovation probability on both diversity and system size, and demonstrate that this effective innovation rule is not only observed in human language but also  emerges as a generating mechanism in modern large language models.

We then show that the same macroscopic laws arise naturally within sampling models with latent structure and broad component-frequency distributions, without
explicitly assuming an innovation mechanism. In this framework, the  realizations are assembled following different rules depending on latent variables, such as the topic composition in books or the taxonomic group in genomes. In contrast to growth models, the latent-variable framework reproduces the same behavior under more general conditions, without requiring a specific asymptotic innovation rule. Finally, we examine data from empirical systems with known hierarchical or thematic organization leading to a latent structure, to assess the relative plausibility of the two explanatory frameworks.
These results provide a clearer interpretation of Heaps' law and its fluctuations in component systems.

\section{Datasets}
\label{sec:data}

All the described datasets can be found in the repository associated with the paper: \url{https://github.com/amazzoli/Diversity_Growth_Topics/}.

\subsection{Books from Project Gutenberg}

This dataset is composed of $3034$ books taken from the Gutenberg database \cite{gutenberg}.
We used the parsed dataset provided by \cite{lahiri2014complexity}, where the metadata, the license information and the transcriber's notes have been removed.
Words were identified by separating them with whitespace or non-alphanumeric characters.
We also converted all uppercase letters to lowercase, and removed all characters other than lowercase ASCII characters (such as numbers and punctuation).

\subsection{Wikipedia}

We analyzed articles from the English Wikipedia, downloaded from the Wikipedia dumps \cite{wikiDumps}.
To parse the raw articles, we used the software provided by \cite{attardiWiki}, which generates XML files that are processed following the same procedure described above for the books. 
The analysis was performed on a random subset of 50000 articles containing more than 250 words.

\subsection{LEGO}

The brick composition of several LEGO sets can be freely downloaded from \cite{rebrickable}.
The basic components are bricks of a given shape and color.
We considered sets with at least $100$ total bricks and at least $20$ distinct brick types, obtaining a total of $7137$ sets.
The dataset also provides a thematic classification of the sets.

\subsection{Proteomics}

The proteome dataset considers genomes in terms of families of protein domains.
Protein domains are basic modular units of proteins \cite{Orengo2005}, which can fold independently and are thermodynamically stable.  
These domains can be grouped into families according to functional and evolutionary similarities.
Different domains belonging to the same family within a genome can occur either in the same protein or in different proteins.
We downloaded all bacterial reference proteomes from UniProtKB (Release 2025\_03) \cite{ahmad2025uniprotkb}. Protein domains within each proteome were identified and mapped to their corresponding family using the Pfam database \cite{paysanlafosse2024pfam}. Each proteome was also mapped to a taxonomic clade according to the NCBI Taxonomy database \cite{schoch2020ncbitaxonomy}, providing a multi-level hierarchical classification. After filtering for genomes containing more than 300 protein domains, the final dataset comprises 8337 bacterial genomes spanning 14693 protein domain families.

\subsection{Large language models}

Section~\ref{sec:DGD_data} uses texts generated by large language models, with input prompts obtained through a random sampling procedure. Specifically, prompts were generated by randomly sampling words from an English-language corpus extracted from Project Gutenberg, using tools provided by the NLTK library~\cite{bird2009nltk}.

All experiments with large language models were carried out using the Hugging Face \texttt{transformers} library \cite{wolf2020transformers}. We employed the pretrained models \texttt{openai-community/gpt-2} \cite{radford2019language} and \texttt{meta-Llama/Meta-Llama-3-8B-Instruct} \cite{touvron2023Llama2openfoundation,llama3modelcard}.

\section{Methods: generative models for  diversity scaling 
}

\subsection{Growth models based on innovation and duplication}
\label{sec:growth_models}

\subsubsection{General definition}

Innovation-duplication models have been widely  used in the complex systems literature to  reproduce emergent empirical statistical features from simple  rules, e.g.,  \cite{zanette2005dynamics, gerlach2013stochastic, karev2002birth, angelini2010mean, rosanova2017modelling}.
These models are based on the idea that a system realization grows by adding components with two possible moves: a duplication of an already-present component, or the  discovery of a new one.
Therefore, a generic realization of a component system is defined by a sequence of \textit{tokens} $(x_1, x_2, \ldots, x_m)$, where each element $x_k$ belongs to a set of unique \textit{components} or \textit{types}, $x_k \in \lbrace c_1, c_2, \ldots c_h \rbrace$.
We call $m$ the object size and $h$ the vocabulary size.
For example, in this framework a book is a sequence of words/tokens, with types representing the words composing its vocabulary.

An innovation-duplication model adds a token at every time step:
\begin{equation*}
(x_1, \ldots, x_{m-1}) \rightarrow (x_1, \ldots, x_{m-1}, x_{m}) ,
\end{equation*}
where, by definition, time coincides with the size $m$.
Two types of moves are possible.
One is an \textit{innovation} event, which occurs with probability $\pn$, that adds a token $x_m$ whose type is still not present in the set of types at time $m-1$.
As a result the vocabulary increases by one unit.
The second possible move is the \textit{duplication} of an existing type $i$,  with probability $p_{old}^i$.
Since no other moves are allowed, $\pn + \sum_i p_{old}^i = 1$.

Models of this class have been used to reproduce the power-law behavior of  component frequencies within a realization \cite{yule1925ii, simon1955class, pitman1996combinatorial, zanette2005dynamics, lagomarsino2009universal}  by introducing a  \quotes{preferential attachment} mechanism in the duplication probability, i.e.,  $p_{old}^i$ increases with the type abundance $n_i = \sum_k \delta_{x_k,c_i}$.
In this work, we focus instead on the innovation dynamics, investigating  the statistics of the number of types as a function of the realization size, $h(m)$, which is known as  Heaps' law in linguistics.

How the \textit{vocabulary}, $h(m)$,  of a realization grows with its size, $m$, is determined by the innovation probability of the underlying  stochastic process, $\pn(h(m), m)$, which can  depend on both the size and the current vocabulary of the realization.
Given the innovation probability,  we can  define the evolution of the stochastic variable $h$ as 
\begin{equation*}
h(m+1) = h(m) + B(h(m), m) ,
\end{equation*}
where $B(h, m) $ is a Bernoulli variable that is $1$ with probability $\pn(h, m)$ and $0$ otherwise.
For the following analysis, it is convenient to consider the continuous-time evolution of the average value:
\begin{equation}
\frac{d \avh}{d m} = \langle \pn \rangle ,
\label{eq:av_h}
\end{equation}
where, for simplicity, we omitted the dependence on $h$ and $m$.
An analogous equation can be defined for the variance, by first computing the evolution of $\langle h^2 \rangle$ and $\avh^2$ from the stochastic equation above, and then by taking again the continuous-time limit to finally obtain the equation for  $\sigma_h^2 = \langle h^2 \rangle - \avh^2$ as
\begin{equation}
\frac{d \sigma_h^2}{d m} = 2 \left( \langle h \pn \rangle - \avh \langle \pn \rangle \right) + \langle \pn \rangle (1 - \langle \pn \rangle) .
\label{eq:var_h}
\end{equation}

\subsubsection{Our case study}
We consider a general innovation-growth framework that, as we will discuss,  subsumes most models of this class proposed in the literature. Given an ensemble of $R$ realizations of the  innovation process, we consider a general innovation probability for realization $j$ as 
\begin{equation}
    p_n(h, m, j) = \alpha_j \frac{h^\beta}{m^\gamma}, 
    \label{eq:p_new_generic}
\end{equation}
which, as a probability,  is capped at unity. 
The parameters $\beta$ and $\gamma$ are non-negative real values: $\beta$ sets the dependence of the innovation rate on the current diversity of the system, while $\gamma$ on its size.  
The parameter $0 < \alpha_j \le 1$ is a realization-dependent prefactor that  introduces heterogeneity in the growth process.
For example, books can belong to different topics or authors with specific  innovation rules. Genomes belong to different taxonomic branches which could evolve with a different dynamics.
The special case $\alpha_j=\alpha$ defines the homogeneous innovation model. 
The numerical simulation of this process is described in Sec. S1 of the Supplementary Material, together with a reliable approximation that significantly speeds up the computation.

\subsection{Sampling models}
\label{sec:sampling}

\subsubsection{Multinomial model}

An alternative framework to study the statistics of diversity $h(m)$ is based on a sampling process that assumes a certain component frequency distribution $\lbrace f_1, f_2, \ldots, f_N \rbrace$, where $N$ is the total number of components in the \quotes{universe},  and $\sum_{i=1}^N f_i = 1$.
The frequency rank distribution is typically chosen as a power-law function to match empirical distributions, i.e.,  $f_{i} \sim r_i^{-\mu}$, where $r_i$ is the rank of the type $i$. This is the classic Zipf's law. 
A synthetic ensemble of realizations can then be generated with a multinomial  sampling process  using the resulting component frequencies   as probabilities. 
The resulting  realizations have marginal statistics that are a \quotes{null} consequence of the  power-law frequencies.

In particular, to study the emerging vocabulary, we define $X_i(m)$ as the Bernoulli variable of sampling at least once the component $i$ after $m$ draws, which has probability $q_i(m) = 1 - (1-f_i)^m$.
The stochastic variable $h$ then reads:
\begin{equation*}
    h(m) = \sum_{i=1}^N X_i(m) .
\end{equation*}
Its average, $\langle h(m) \rangle = \sum_{i=1} q_i(m)$, is determined by the underlying frequency distribution, leading to the relation $\langle h(m) \rangle \approx m^{1/\mu}$ \cite{van2005formal, lu2010zipf, eliazar2011growth} for power-law frequencies $f_{i} \sim r_i^{-\mu}$ with $\mu > 1$, in the limits $m \gg 1$, $N \gg 1$ and $m \ll N^\mu$ (vocabulary far from saturation).

In a similar way, we can also compute the variance:
\begin{equation}
     \sigma_h^2(m) = \sum_{i=1}^N q_i(m) [1 - q_i(m)] \le \sum_{i=1}^N q_i(m) = \langle h(m) \rangle,
     \label{eq:var_sampling}
\end{equation}
which is always smaller than or equal to the average since $0 \le q_i(m) \le 1$ and, therefore, bounded by the Poisson scaling $\sigma_h^2 = \avh$.

\subsubsection{Sampling with latent variables}

A richer extension of the sampling model considers the presence of latent variables (or ``topics''), such as in the famous Latent Dirichlet Allocation model \cite{blei2003latent}.
Assuming the presence of $T$ topics, each topic $t$ will have a specific distribution of type frequencies, $g_i(t)$.
A given realization $j$ is defined by a mixture of topics  by $P(t|j)$, and it is built by sampling each type with a two-step process: first we sample the topic from $P(t|j)$ and, second, we sample the type from the topic-specific probabilities $g_i(t)$.
In the large limit, this process generates a set of realizations with frequencies $f_i(j) = \sum_t P(t|j) g_i(t)$.
Differently from the simple multinomial case, each realization can have a specific component distribution, introducing a crucial additional layer of variability.
This additional variability  can lead to a superlinear scaling of the variance of the vocabulary $h$ with its  average \cite{altmann2016statistical}, i.e.,  $\sigma_h^2 > \avh$, breaking the Poisson bound of Eq. \ref{eq:var_sampling}.

\subsubsection{Our case study}
\label{sec:sampling_exe}

We will analyze a Dirichlet model with $T$ topics, where the probabilities over the topics $P(t|j)$ are sampled from a Dirichlet distribution with all the concentration parameters equal to $\Theta$.
That means that, for $\Theta \rightarrow \infty$, the distribution tends to be uniform over the $T$ topics, while, for $\Theta \rightarrow 0$, it becomes a delta function of one random topic.  
We choose the frequencies at a given topic to have a power-law tail and a \quotes{core} of $c(t)$ types with maximal constant frequency, that, ideally, represent the topic-defining types:
\begin{equation}
g_i(t) \propto \left\{
    \begin{aligned}
    & 1 & \text{for } \; r_i(t) \le c(t)
    \\
    &1/(r_i(t) - c(t))^\gamma & r_i(t) > c(t)  
    \end{aligned} .
\right.
\label{eq:LDA_freqs}
\end{equation}
For each topic, the component ranking $r_i(t)$ is a random permutation of the type indexes, and the number of core components, $c(t)$, is sampled from a gamma distribution with typical values much smaller than the total vocabulary $N$.  

\begin{figure}
    \centering
    \includegraphics[width=\linewidth]{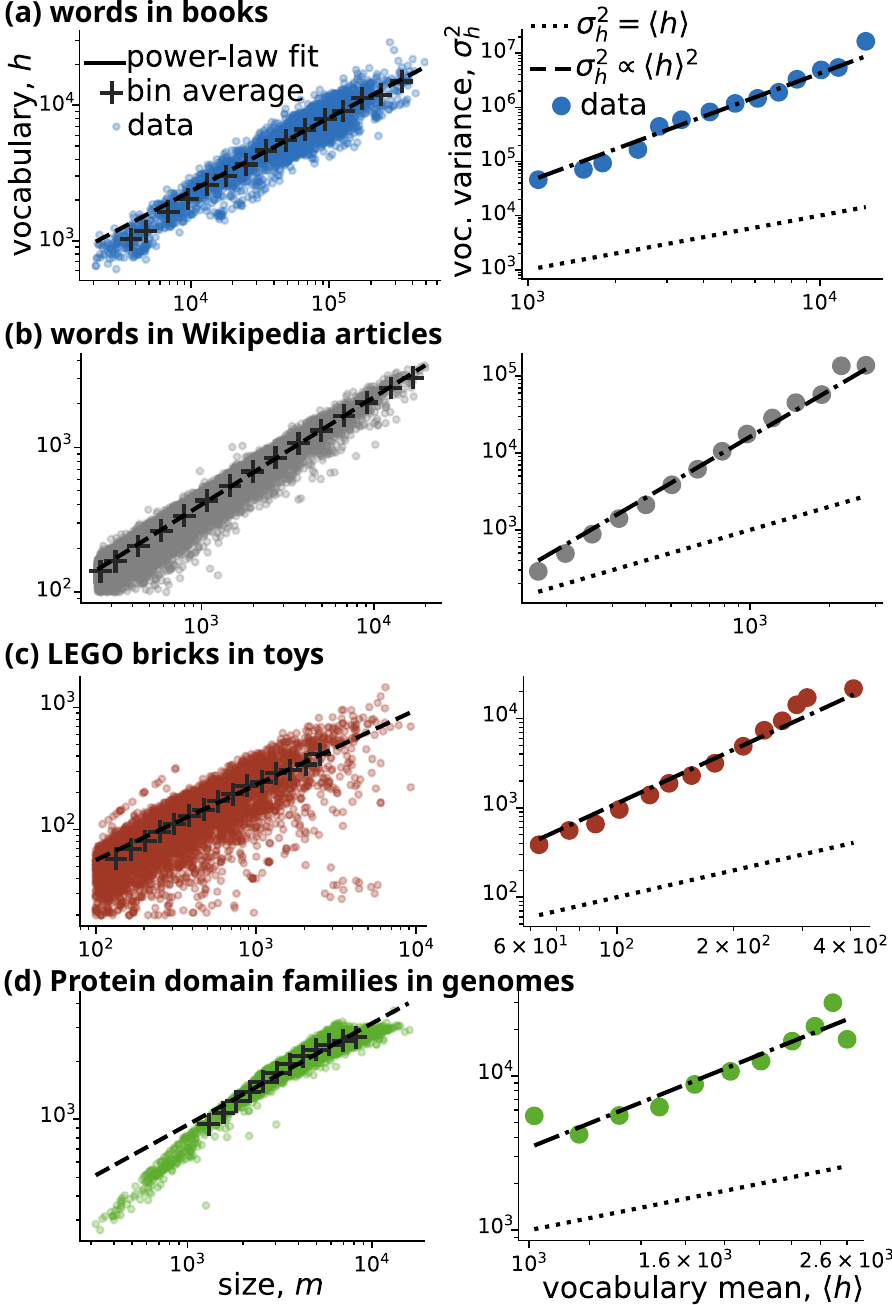}
    \caption{
    \textbf{Heaps' law and quadratic fluctuation scaling of diversity in complex component systems.} The left-hand panels show Heaps' law, i.e., the sublinear dependence of the vocabulary size on systems size for
    (a) books from the Project Gutenberg, (b) articles from Wikipedia, (c) LEGO bricks in LEGO boxes, and (d) protein domain families in genomes (see the dataset section for details).
    The Heaps' law is fitted with a power-law function with exponent: (a) $\nu = 0.54$, (b) $\nu = 0.74$, (c) $\nu = 0.53$ (for sizes larger than $1200$), (d) $\nu = 0.62$.
    The right-hand panels show the scaling of the variance of diversity with the average diversity, divided in logarithmic equally spaced bins. 
    The dotted line is the Poisson prediction $\sigma_h^2 = \avh$, while the dashed line is a quadratic scaling law $\sigma_h^2 = c \avh^2$, where $c$ is obtained with a fit, which agrees with the data.
    }
    \label{fig:heaps_taylor}
\end{figure}

\section{Results}

\subsection{
Heaps' law and its quadratic fluctuation scaling in  component systems}
\label{sec:heaps_taylor}

We start by analyzing the statistics of the number of unique components, the vocabulary $h$, as a function of the total number of components, the size $m$, for systems of radically different nature (texts, LEGO constructions, or proteomes) composed of elementary types/components (words, bricks, or protein families).
Considering all realizations in a given dataset,  a sub-linear trend, called Heaps' law,  can be observed (left plots of Fig. \ref{fig:heaps_taylor}).

Empirically, in all four different datasets the average vocabulary, $\avh(m)$, grows approximately as a power law of size: 
\begin{equation}
\avh \propto m^\nu ,
\label{eq:heaps_data}
\end{equation}
with $0 < \nu < 1$.

The second quantity we are interested in are the vocabulary fluctuations: 
\begin{equation}
\sigma_h^2 = \langle h^2 \rangle - \avh^2 \propto \avh^\rho ,
\label{eq:NSA_data}
\end{equation}
where $\rho$ is the scaling exponent of the variance with the average.
All the different systems show a super-linear scaling compatible with $\rho=2$ (right column of plots of Fig. \ref{fig:heaps_taylor}). This fluctuation scaling  implies that these systems are non-self-averaging, since relative fluctuations do not vanish as the system size diverges.
This fluctuation scaling  deviates from the expectation from a simple sampling process that would naturally predict a Poisson scaling, Eq. \ref{eq:var_sampling}, providing a clear and robust empirical feature that can be used to falsify generative models.

\subsection{Growth models reproduce quadratic fluctuation scaling only under diversity-dependent innovation}
\label{sec:DGD_model}

\subsubsection{Linear diversity-generates-diversity mechanism is sufficient for quadratic fluctuation scaling }

To reproduce both the average sublinear growth, Eq. \ref{eq:heaps_data}, and the non-self-averaging behavior, Eq. \ref{eq:NSA_data}, previous works \cite{tria2018zipf, tria2020taylor} adopted specific innovation-duplication models  introduced in Sec. \ref{sec:growth_models}.
In particular,  models that have been shown to successfully reproduce these two statistical properties  are the Chinese Restaurant Process (CRP) \cite{pitman1997two, bassetti2009statistical}, with $\pn = (\theta + \alpha h)/(\theta + m)$, and a generalization of Pólya's urn \cite{tria2014dynamics} with $\pn = (h_0 + \nu h)/(h_0 + (\nu+1)h + \rho m)$.
Moreover, a specific choice of  parameters of the general model defined in Sec. \ref{sec:growth_models} through Eq. \ref{eq:p_new_generic}, can also reproduce a sublinear average vocabulary growth and a quadratic fluctuation scaling (see Sec. 2 of SM). Specifically, the  model has to be specified with $\beta=\gamma=1$ and a constant $\alpha$, thus with an innovation probability $\pn = \alpha h / m$. 
Crucially, all these models display an innovation probability that becomes $\pn \sim h/m$ in the limit of large sizes.
The probability decreases with size in order to give  rise to the sub-linear vocabulary growth, but,  at the same time,  it becomes larger for more diverse vocabularies at a given size, implementing  a sort of rich-gets-richer  mechanism or, as we call it in this context, a diversity-generates-diversity  mechanism.

Conversely, when $\pn$ does not depend on the vocabulary, one can obtain the sublinear growth, but not a quadratic variance scaling.
This can be seen starting from Eq. \ref{eq:p_new_generic} with a constant $\alpha$, with $\gamma < 1$ and $\beta = 0$, a process that is equivalent to a generalization of the Yule-Simon model \cite{zanette2005dynamics}, $\pn = \alpha \; m^{-\gamma}$.
By integrating Eq. \ref{eq:av_h},  we obtain the sublinear average growth of the vocabulary: $\avh \propto m^{1-\gamma}$.
However, its variance and, in general, the variance of every model with $\pn$ independent of $h$, is determined by Eq. \ref{eq:var_h}, which simplifies to:
\begin{equation*}
\frac{d \sigma_h^2}{d m} = \langle \pn \rangle (1 - \langle \pn \rangle) \le  \langle \pn \rangle = \frac{d \avh}{d m} .
\end{equation*}
The inequality is due to the fact that $0 < \pn < 1$ while the second equality comes from Eq. \ref{eq:av_h}.
Therefore,  the growth of the variance is bounded by the growth of the average. Since the initial conditions are $\avh = 1$ and $\sigma_h^2 = 0$, the fluctuations are thus bounded by the Poisson scaling, $\sigma_h^2 \le \avh$.

In conclusion, the dependence of the innovation rate on the vocabulary in growth models is key to reproducing the two desired scaling behaviors. This result  naturally raises the question of whether functional dependencies other than  $\pn \sim h/m$ can lead to different fluctuation scalings.

\subsubsection{A linear diversity-generates-diversity mechanism is necessary for quadratic fluctuation scaling}

\begin{figure}
    \centering
    \includegraphics[width=\linewidth]{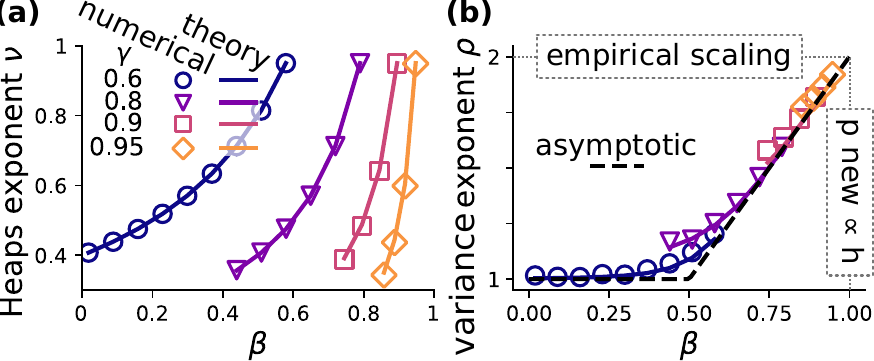}
    \caption{
    \textbf{In growth models,   how the  innovation-rate depends on systems size and on  current diversity  sets the average diversity scaling and its fluctuations.}
    The exponents are computed numerically (symbols) for different parameters $\beta$ (x-axis) and $\gamma$ (colors) of the model in Eq.~\ref{eq:p_new_generic} with constant $\alpha$.
    We chose values of $\gamma < 1$ and $\beta < \gamma$ to be in the region (I) of the parameter space.
    Moreover, we chose a minimum value of $\beta$ such that $\nu = (1-\gamma)/(1-\beta) > 0.35$, in order to reach large values of $\avh \sim m^\nu$, needed for satisfying the asymptotic relations, in a reasonable time.
    Simulations set $\alpha=\nu$ and are run for $m=1.5 \cdot 10^6$ steps, using the approximation of S1 with $\Delta m = 100$.
    The exponents are computed with the discrete logarithm derivative in the last window of $2 \cdot 10^5$ steps and averaged over $R=2 \cdot 10^5$ realizations.
    The lines in (a) are the theoretical expectation from Eq. \ref{eq:av_h_mean_field}.
    The black dashed line of (b) is the asymptotic value of Eq. \ref{eq:var_h_homog} in the limit of infinite average, while the colored lines are the effective exponents for finite $\avh$, the left term of Eq. \ref{eq:var_h_homog}. See Eq. S7 for obtaining the effective exponent from Eq. \ref{eq:var_h_homog}.
    }
    \label{fig:NSA_exp}
\end{figure}

Here,  we consider the general innovation model described by Eq. \ref{eq:p_new_generic}, with generic $\beta$ and $\gamma$, in its homogeneous version with a constant prefactor $\alpha$.
The first problem we want to solve is to identify the range of parameters that can recover the empirical sublinear scaling of the Heaps' law.

To this end, a useful approximation is a \quotes{mean field} version of Eq. \ref{eq:av_h}, where we move the average expectation within the innovation probability $\langle \pn (h,m) \rangle = \pn (\avh, m)$, obtaining the equation
\begin{equation*}
    \frac{d \bar{h}}{d m} = \alpha \frac{ \bar{h}^\beta}{m^\gamma} ,
\end{equation*}
where $\bar{h}$ is the mean-field vocabulary variable.
The specific trajectory that describes the average growth is obtained by solving the equation from the initial condition $\bar{h}(1) = 1$.
Having defined $\nu = (1-\gamma)/(1-\beta)$ and using a prime for this specific trajectory, we can obtain:
\begin{equation}
    \bar{h}'(m) = \left(\frac{\alpha}{\nu}\right)^{1/(1-\beta)} \; m^\nu .
\label{eq:av_h_mean_field}
\end{equation}
To be consistent with the empirical growth, we need $0 < \nu < 1$, constraining the  range of parameters to region (I): $\gamma > \beta$ for $\beta < 1$ and $\gamma < 1$; or (II): $\gamma < \beta$ for $\beta > 1$ and $\gamma > 1$.

While the region (I) reproduces the desired sublinear trend (Fig. S2a and Fig. S2bI), in the region (II) the simulated trajectories show an approximately bimodal behavior, since they  accumulate either on the upper bound $h = m$ or do not grow $h = O(1)$ (Fig. S2bII).
Moreover, while the numerical exponent of (I) matches the mean field prediction (Fig. \ref{fig:NSA_exp}a), in (II) the exponent is $1$ (Fig. S2a).
This anomalous bimodal behavior is also highlighted by the kurtosis Fig. S3, which shows a strong deviation from a uni-modal Gaussian behavior.

To mathematically understand the difference between the two regions, we study the stability of the average trajectory in Eq.\ref{eq:av_h_mean_field}.
Specifically, we define a new variable by rescaling the vocabulary by the average growth $x = \bar{h} / \bar{h}'$ and we consider its dynamics:
\begin{equation}
    \frac{d x}{dm} = \frac{\nu}{m} \left( x^\beta - x \right) .
    \label{eq:rescaled_h}
\end{equation}
The value $x=1$, corresponding to $\bar{h} = \bar{h}'$, is a fixed point, which is stable only for $\beta < 1$,  region (I), implying that fluctuations lead back to the average growth $\bar{h}'$.
For $\beta > 1$, region (II), the trajectories are unstable, deviating from the sub-linear average growth towards the boundaries.
Eventually, the trajectories that grow linearly will dominate the average, leading to $\avh \sim m$, and breaking the mean field prediction.
In general, the stability analysis of  Eq. \ref{eq:rescaled_h}  recapitulates the qualitative behavior of the average growth in all the parameter space $\beta$, $\gamma$, as shown in Fig. S2c.

In conclusion, the empirical  power-law growth of the vocabulary is reproduced only in region (I) of the parameter space, and we can now focus on the fluctuation scaling.
To make analytical progress,  we need to assume that the innovation probability, Eq. \ref{eq:p_new_generic}, is approximately linear in $h$ around $\avh$ and within the range $\avh \pm \sigma_h$.
As we show in Sec. S3, this allows us to explicitly compute $\langle \pn \rangle$ and $\langle h \pn \rangle - \avh \langle \pn \rangle$, and solve Eq. \ref{eq:var_h}:
\begin{equation}
\sigma_h^2 = \frac{\avh - \avh^{2 \beta}}{1 - 2\beta} \xrightarrow[\avh \rightarrow \infty]{}
\begin{cases}
\avh^{2 \beta} \;\; \text{for} \;\, \beta > 1/2 \\
\avh \;\; \text{for} \;\, \beta < 1/2 .
\end{cases}
\label{eq:var_h_homog}
\end{equation}
This expression is tested in Fig. \ref{fig:NSA_exp}b, showing that growth models can  recover all possible  exponents between $1$ and $2$ by properly tuning $\beta$.
Notably, the asymptotic expression for large system sizes does not depend on $\gamma$, which only determines how quickly the large-$\avh$ limit is reached through the value of $\nu$.
The empirical quadratic fluctuation scaling is reached only for $\beta \rightarrow 1$, which also implies $\gamma \rightarrow 1$ to remain in region (I) of the parameter space. 
Therefore, the only  form of innovation probability that can jointly recover the average power-law growth of Heaps' law and its quadratic fluctuation scaling  is the linear diversity-generates-diversity mechanism $\pn \sim h/m$.
While this linear asymptotic dependence is a necessary condition for the quadratic scaling, models can have additive terms in the innovation probability. 
To understand  their effect , we study the illustrative example of  Chinese-Restaurant-Process-like models, in which  $\pn = (\theta + \alpha h^\beta) / (\theta + m^\gamma)$.
Fig. S4 shows that the Heaps and the fluctuation scaling exponents still follow Eqs. \ref{eq:av_h_mean_field} and \ref{eq:var_h_homog}, while deviations can be observed only for very large values of $\theta$.
More precisely, the results hold for $\theta \ll \alpha \avh(m)^\beta$ and $\theta \ll m^\gamma$, with $m$ being the size at which the exponents are computed. Therefore,  our conclusions can be extended to all  models whose innovation probability behaves as Eq. \ref{eq:p_new_generic} in the regime of large system sizes.

\subsection{Diversity-generates-diversity is an emergent property of large language models}
\label{sec:DGD_data}

\begin{figure}
    \centering
    \includegraphics[width=\linewidth]{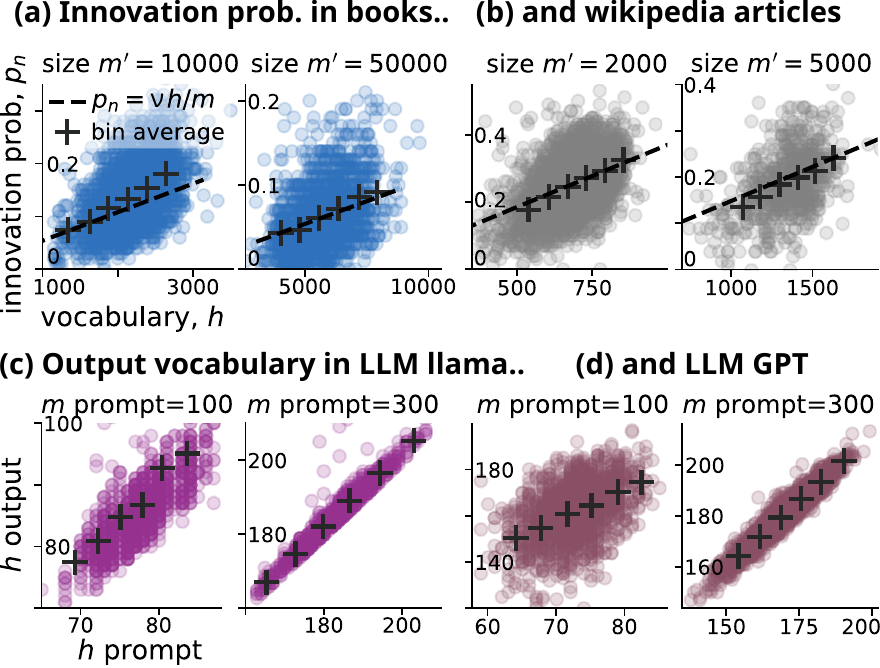}
    \caption{
    \textbf{Future vocabulary diversity depends linearly on current diversity in texts generated by both humans and large language models.} Panels
    (a) and (b) show the empirical innovation probability after the first $m'$ words: $\pn = (h(m'+\Delta m) - h(m')/\Delta m$ ($\Delta m = 100$) in human texts.
    This probability was was estimated across different texts of the Gutenberg project (a) or Wikipedia (b) and plotted against $h(m')$.
    The gray crosses are the binned average, while the dashed line is the prediction of $\pn=\nu h / m'$, with $\nu$ being the power-law exponent fitted from the average Heaps (values in Fig. \ref{fig:heaps_taylor}). Panels
    (c) and (d) show the vocabulary of the output text of a Large Language Model as a function of the vocabulary of the input/prompt for different numbers of words in the prompt $m$ and for two models: Llama (c) and GPT (d).
    Input prompts were generated randomly as described in the Dataset section.
    Outputs are cut at $m_{\rm output} = 200$ and $400$ words for the two prompt sizes of (c) and at $m_{\rm output} = 330$ words for both the prompts of (d).
    These values of $m_{\rm output}$ are chosen to be slightly smaller than the typical generated output size.
    }
    \label{fig:p_new_data}
\end{figure}

The presence of an effective diversity-generates-diversity mechanism can be  tested in systems that have a natural temporal ordering,  and thus in which vocabulary trajectories can be defined.
Texts are a paradigmatic example, because we can count how many different words are present in the first $m'$ words, defining a trajectory $h(m')$ for each realization.
The effective innovation probability can then be computed as a numerical derivative with respect to $m'$, as reported in 
Fig. \ref{fig:p_new_data}a,b for texts at different  fixed sizes $m'$.
As expected from our theory, the average innovation probability is linear in the vocabulary, showing that books with a  richer vocabulary at a given size will further increase their word diversity in the near future.
The shape of the innovation probability is compatible with  $\pn = \alpha h / m$, with the prefactor $\alpha$ set by the power- law exponent $\nu$ of the average Heaps, as predicted by the model. 
While this is a consistency check for our analytical theory, the growth model should be realistically interpreted  only as an effective description since  language production by humans is much more  complex  than a word-by-word progressive addition. 

However, the generative process of Large Language Models (LLMs) is closer to the model description,  as they sample from a next-word probability distribution given the current context. Therefore, we can check if a diversity-generates-diversity mechanism is inherently implemented in LLM text production  by limiting confounding factors such as semantics, topicality and style. To this aim, we built  textual prompts of fixed size $m_{\rm prompt}$ by randomly sampling words from the empirical word frequency distribution, as described in more detail in the Methods section.
These prompts have some variability in the vocabulary size $h_{\rm prompt}$ because of random fluctuations, but otherwise they do not have any particular meaning or structure.
The LLM will generate an output text following the prompt, which we cut at a given size $m_{\rm output}$.
Interestingly, the number of different words in the output, $h_{\rm output}$, strongly correlates with $h_{\rm prompt}$ and shows an average linear relation (Fig. \ref{fig:p_new_data}c,d) as expected from a linear diversity-generates-diversity mechanism. 
Notice that the y-axis is not exactly the innovation probability, but $h_{\rm output} / m_{\rm output}$ tends to it in the limit $m_{\rm output} \rightarrow 0$.

\subsection{Topicality  robustly leads to a quadratic scaling of  diversity fluctuations }
\label{sec:NSA_topic}

\subsubsection{Fluctuation scaling from topicality}

We now turn to an alternative modelling framework based on latent variables. Unlike growth models, which we have shown require a specific innovation rule to reproduce quadratic fluctuation scaling, latent-variable models can robustly generate the same behavior through heterogeneity across realizations, without requiring any specific innovation dynamics.

 A simple sampling process from a Zipf-like component frequency distribution is known to reproduce the average Heaps' law~\cite{mazzolini2018heaps}, but it cannot generate quadratic fluctuation scaling because independent sampling naturally produces Poisson-like fluctuations. By contrast, latent variables introduce additional variability across realizations, allowing the diversity variance to grow superlinearly~\cite{gerlach2013stochastic}.

In linguistics, latent variables are naturally interpreted as topics, but their meaning depends on the system under consideration. More generally, they represent hidden sources of heterogeneity that shape component frequencies. For example, different biological functions require different repertoires and abundances of protein families in genomes, while different LEGO themes favor different subsets of bricks.

We consider a general latent-variable model and derive the conditions under which it reproduces quadratic fluctuation scaling.
Specifically, we generically assume that the vocabulary statistics depend on latent variables $t$, such that $P(h) = P(h|t)P(t)$.
We further assume that the generative process satisfies Heaps' law. A simple possibility would be that  the average vocabulary conditioned on  the topic $\mathbb{E}_{h|t}[h] = \sum_h h P(h|t)$ is given by 
\begin{equation}
    \mathbb{E}_{h|t}[h] = k(t) \; m^\nu . 
    \label{eq:avh_at_topic}
\end{equation}
In this case,  all topics induce the same Heaps exponent $\nu$, while the prefactor $k(t)$ depends on the latent variable.

Given Eq. \ref{eq:avh_at_topic}, the  average vocabulary across realizations can be obtained by averaging over  topics as
\begin{equation}
    \avh \coloneqq \mathbb{E}_{h}[h] = \mathbb{E}_{t}[\mathbb{E}_{h|t}[h]] = \mathbb{E}_{t}[ k(t) ] m^\nu ,
    \label{eq:avh_cond}
\end{equation}
which retains the same scaling as the topic-conditioned averages.

We can now decompose the vocabulary variance  using the law of total variance:
\begin{equation}
    \begin{aligned}
    \sigma^2_h \coloneqq \text{Var}_{h}[h] = & \;\mathbb{E}_{t}\left[\text{Var}_{h|t}[h] \right] + \text{Var}_t \left[ \mathbb{E}_{h|t}[h] \right] = \\
    & \;\mathbb{E}_{t}\left[\text{Var}_{h|t}[h] \right] + \text{CV}_t^2[k(t)] \avh^2,
    \end{aligned}
    \label{eq:var_cond}
\end{equation}
where the second line uses Eq. \ref{eq:avh_at_topic} and \ref{eq:avh_cond}, and the coefficient of variation (CV) is defined as $\text{CV}_t^2[k(t)] = \text{Var}_t \left[ k(t) \right] / \mathbb{E}_t \left[ k(t) \right]^2$.
This decomposition separates two distinct sources of fluctuations.
The first term describes the variability given a fixed topic, which in general is expected to remain sub-Poissonian (see also the two examples below).
The second term instead reflects the variability induced by topics. Since this contribution scales quadratically with the average vocabulary, it eventually dominates for sufficiently large $\avh$, naturally giving rise to the quadratic fluctuation scaling we observe empirically. The amount of topic-dependent variability, quantified by $\mathrm{CV}_t[k(t)]$, determines the scale of the crossover to this asymptotic regime.


In Sec. S4, we extend this analysis to the case in which the topic dependence is contained in the exponent, $\nu(t)$, rather than in the prefactor $k(t)$. For sufficiently small exponent variability, quadratic fluctuation scaling is preserved up to a logarithmic correction.
In practice, this correction only slightly modifies the effective exponent and is expected to be difficult to distinguish from a pure quadratic law with the limited number  of data points of empirical datasets and over the  finite range of $\avh$  accessible.


Unlike the growth framework, the model presented in this section does not jointly explain both Heaps' law and quadratic fluctuation scaling. Instead, it assumes that Heaps' law holds for each latent variable and shows that quadratic fluctuation scaling follows naturally from heterogeneity across realizations through the law of total variance. Consequently, any mechanism that reproduces the average Heaps' law automatically acquires quadratic fluctuation scaling once latent variability is introduced.

\subsubsection{The phenomenological equivalence between topicality and a linear  diversity-generates-diversity mechanism}

Within this framework, we can ask what effective  innovation dynamics would be inferred from the observed vocabulary growth. The innovation probability can be extracted from data as the derivative of the vocabulary $h(m)$ with respect to the system size $m$. More precisely, we can compute the average innovation probability conditioned on a latent variable as
\begin{equation}
    \mathbb{E}_{h|t}[\pn] = \frac{d \mathbb{E}_{h|t}[h]}{d m} = \nu k(t) m^{\nu - 1} = \nu \frac{\mathbb{E}_{h|t}[h]}{m} ,
    \label{eq:pnew_topic}
\end{equation}
where we used the sublinear size dependence of Eq. \ref{eq:avh_at_topic}.
Therefore, realizations with a larger vocabulary at a fixed size also exhibit a larger effective innovation probability.
Remarkably, this is exactly the same linear dependence on diversity that we found to be necessary in the growth framework. Here, however, it does not arise from an explicit innovation mechanism. Instead, it emerges as a consequence of the latent heterogeneity that influences the component statistics.

This effective diversity-generates-diversity mechanism becomes observable when the variability of the average vocabulary across latent variables $\text{Var}_t \left[ \mathbb{E}_{h|t}[h] \right]$ exceeds the typical fluctuations within a given latent variable $\mathbb{E}_{t}\left[\text{Var}_{h|t}[h] \right]$. These two contributions are precisely the two terms appearing in the law of total variance in Eq. \ref{eq:var_cond}, with the term $\text{Var}_t \left[ \mathbb{E}_{h|t}[h] \right]$ dominating for sufficiently large $\avh$. 
Consequently, in the same regime where the quadratic fluctuation scaling emerges, the effective inferred innovation probability  approaches $p_n \sim h/m$.

Sec. S4 extends these results  to the case of a topic-dependent Heaps exponent $\nu(t)$ and shows that the quadratic scaling is preserved, with a logarithmic correction with respect to Eq. \ref{eq:pnew_topic}.

In conclusion, we showed that the presence of latent variables  not only asymptotically produces quadratic fluctuation scaling,  but also induces the same effective innovation law (i.e., $\pn \sim h/m$) necessary in growth models  to explain diversity fluctuations. As a consequence, the two descriptions become phenomenologically indistinguishable if one observes only the average Heaps' law, the fluctuation scaling, and  the effective innovation probability inferred from trajectories. 
The following examples illustrate this correspondence in two concrete generative models in more detail.

\begin{figure}
    \centering
    \includegraphics[width=\linewidth]{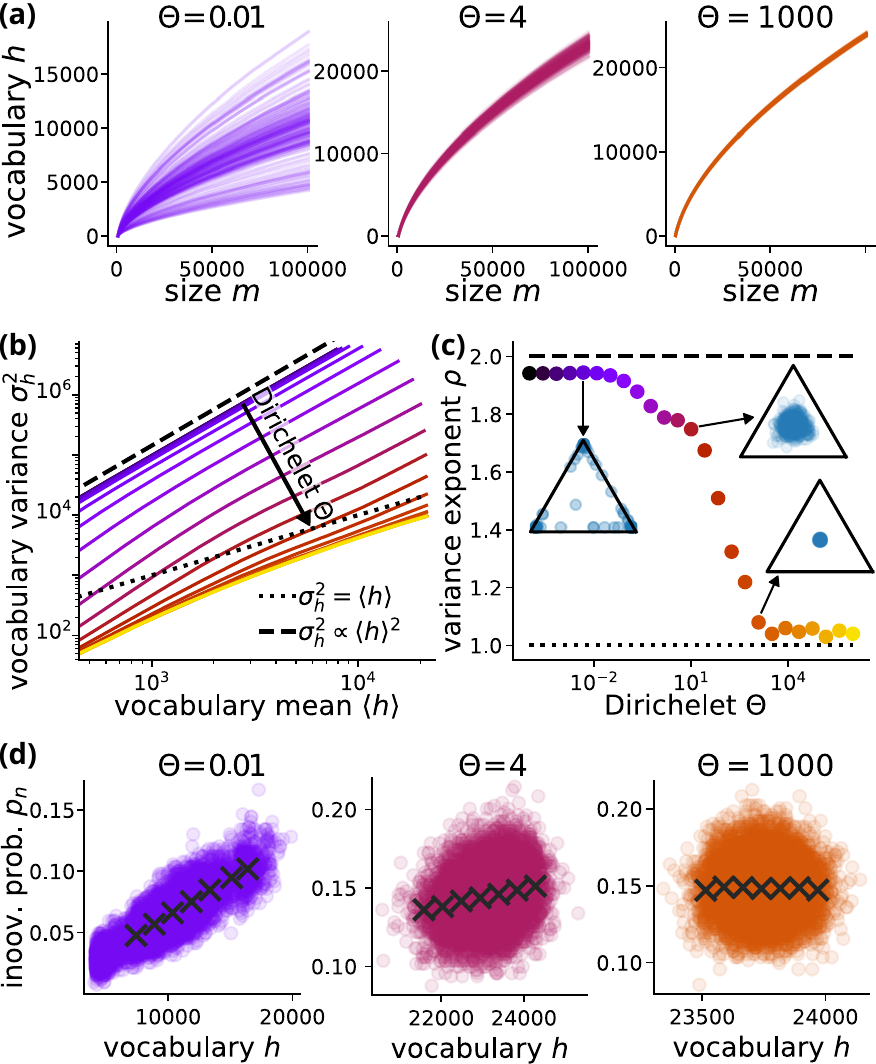}
    \caption{
    \textbf{A sampling model with topics transitions from  Poisson- to quadratic  diversity fluctuations for a sufficient level of topic heterogeneity.}
    (a) Vocabulary trajectories as a function of size in the topic model of Sec. \ref{sec:sampling_exe} for different values of the Dirichlet concentration parameter $\Theta$.
    The model considers $T=20$ topics with a number of core components sampled from a Gamma distribution with average $50$ and standard deviation $50$.
    The power-law exponent of the frequency distributions is $\mu = 1.5$, the trajectories are run for $m_{max} = 10^5$ steps, and the total number of types is $N = 100 \, m_{max}^{1/\mu}$ in such a way that trajectories are far from saturation.
    (b) Vocabulary variance as a function of the average for different values of $\Theta$ (different colors), computed with $5 \cdot 10^5$ realizations.
    We explicitly report  the quadratic and the  Poisson scaling as  black lines for comparison.
    (c) Numerical evaluation of the variance-average scaling exponents fitted with a least squares method at the end of the trajectories in a bin $\Delta \avh = 0.15 \avh(m_{max})$.
    For three values of $\Theta$, we show what a simplified 3-topic simplex looks like with points sampled from a Dirichlet distribution with that $\Theta$.
    (d) Numerical evaluation of the innovation probability for different values of $\Theta$ at the end of the trajectories. $\pn = \Delta h / \Delta m$ with $\Delta m = 500$.
    }
    \label{fig:LDA_NSA}
\end{figure}

\subsubsection{Example 1: Latent-Dirichlet-Allocation sampling model}

As an illustrative example of a sampling model with latent variables, we consider the simple model previously introduced in  Sec. \ref{sec:sampling_exe} and inspired by the classic Latent Dirichlet Allocation (LDA) topic model~\cite{blei2003latent}. 
In this model,  each topic $t$ has a specific frequency distribution $g_i(t)$, with a core of components with constant maximal frequency followed by a power-law tail (Eq. \ref{eq:LDA_freqs}).
The power-law tail is known to generate a sublinear average growth of Heaps' law \cite{van2005formal, lu2010zipf, eliazar2011growth,mazzolini2018heaps}, satisfying the hypothesis of Eq. \ref{eq:avh_at_topic}.
In this model,  the amount of variability generated by topics can be controlled through the parameter $\Theta$ of the Dirichlet distribution used for generating the topic composition, $P(t|j)$.
For $\Theta \rightarrow 0$, the Dirichlet will generate delta distributions with the peak in a randomly chosen topic.
This creates single-topic realizations having a frequency distribution that matches the one of its topic $g_i(t)$, implying topic-specific Heaps' laws.
Fig. 4a shows this scenario for $\Theta = 0.01$, where we can observe separate bundles of trajectories belonging to separate topics.
This is the regime of maximal vocabulary variability, and we expect a quadratic fluctuation scaling. 
Conversely, for $\Theta \rightarrow \infty$, realizations will have the same flat topic composition, $P(t|j) = 1/T$ (for $T$ topics), and, eventually, the same frequency distribution: $f_i(j) = \sum_t P(t|j)g_i(t) = \sum_t g_i(t) / T$.
As a result, this scenario becomes equivalent to the simple multinomial sampling with diversity  variability bounded by a Poisson scaling (Eq. \ref{eq:var_sampling}), Fig. \ref{fig:LDA_NSA}a for $\Theta = 1000$.

Fig. \ref{fig:LDA_NSA}a,b,c show the transition between a quadratic scaling and the Poisson regime  as a function of $\Theta$.
While model details can influence the specifics, the general behavior can be interpreted as a balance between the two terms of Eq. \ref{eq:var_cond}. 
Specifically, the coefficient in front of the quadratic term is determined by the amount of topic variability, which vanishes in the limit $\Theta \rightarrow \infty$, but becomes increasingly relevant for smaller values of $\Theta$. Again for large system sizes a quadratic fluctuation scaling is expected as long as some latent heterogeneity is present, but for small system sizes deviations are possible in the crossover region of topic variability.    
Consistently, the  numerical effective innovation probability is linear with $h$ as soon as  $\Theta$ is small enough, transitioning to an almost flat dependency for larger values, Fig. \ref{fig:LDA_NSA}d.

\subsubsection{Example 2: Growth model with intrinsic heterogeneity}

As a second example, we consider an innovation growth model defined by Eq. \ref{eq:p_new_generic},  where the prefactor $\alpha_j$ is variable across realizations, capturing a realization-specific topic composition. 
For simplicity, we consider the case $\beta = 0$, having then $\pn = \alpha_j m^{-\gamma}$.
If $\alpha$ is constant, this setting cannot reproduce the quadratic fluctuation scaling, Eq. \ref{eq:var_h_homog}, for $\beta=0$.
However, if $\alpha_j$ changes with realizations,  it plays the role of a latent variable: $P(h) = P(h|\alpha)P(\alpha)$.
The average vocabulary  growth conditioned on a fixed $\alpha$ is a power law (Eq. \ref{eq:av_h_mean_field}), thus  satisfying the  conditions discussed in Sec. \ref{sec:NSA_topic}.
The variance of the process is then given by Eq. \ref{eq:var_cond}, where the first term can be obtained from Eq. \ref{eq:var_h_homog} and the second term by Eq. \ref{eq:av_h_mean_field}:
\begin{equation}
    \sigma^2_h \approx \avh + \text{CV}^2[\alpha] \avh^2 ,
    \label{eq:varh_hetero}
\end{equation}
Consistently, in Sec. S5 we obtain the same result by directly solving the model variance through Eq. \ref{eq:var_h}.
The parameter that controls the asymptotic limit is the coefficient of variation of $\alpha$, with a quadratic fluctuation scaling emerging for $\avh \gg 1 / \text{CV}^2[\alpha]$, as tested in detail in Fig. S5.

In Sec. S5, we also study the full model for $\beta > 0$ and Fig. S6 shows numerically that, by adding topic variability, the model converges to a quadratic fluctuation scaling for every value of $\beta$ and $\gamma$.

\begin{figure}
    \centering
    \includegraphics[width=\linewidth]{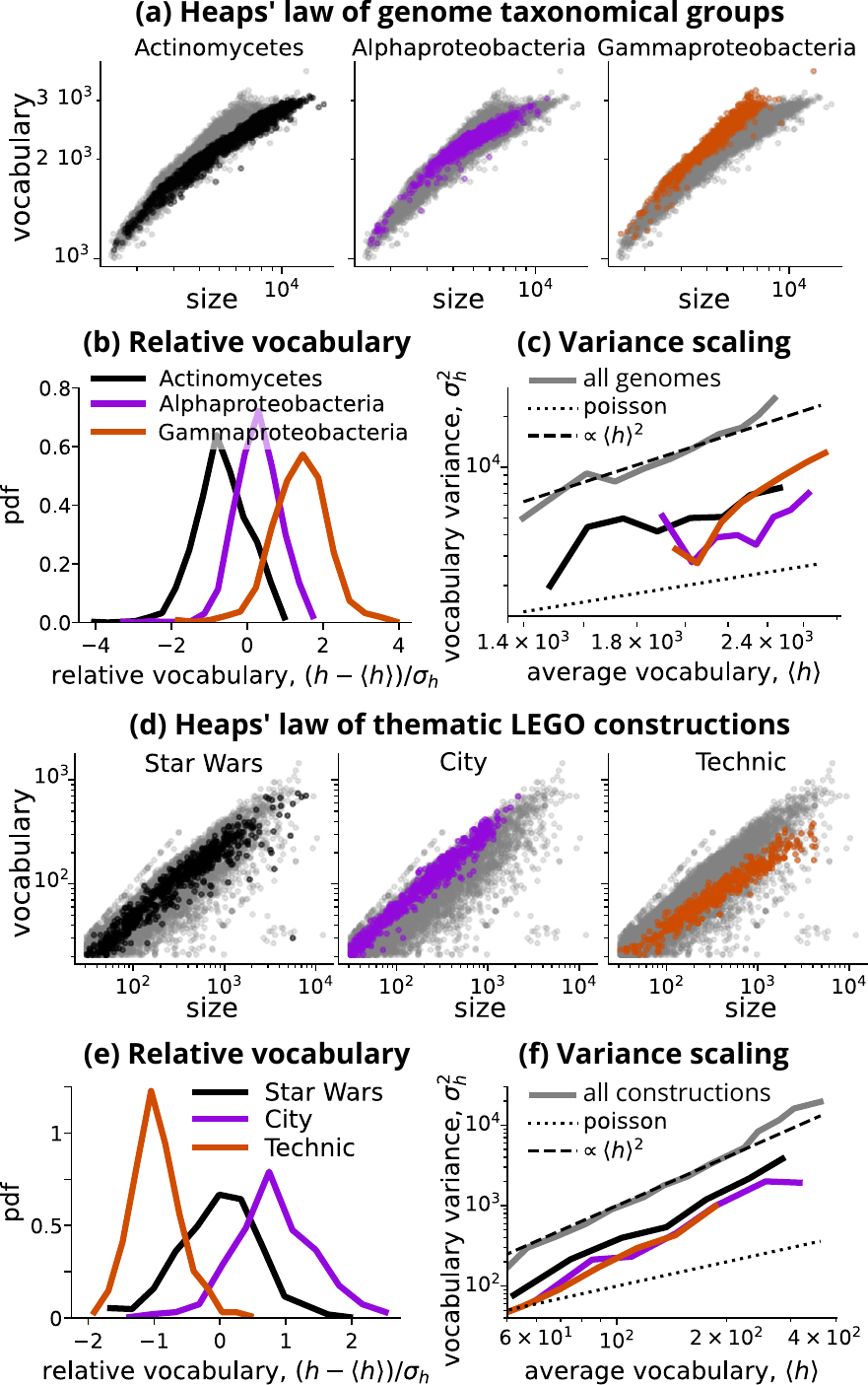}
    \caption{
    \textbf{Empirical component systems where topics are partially known show topic-dependent vocabulary diversity scalings, associated with non-self-averaging Heaps' laws.}
    Ensemble Heaps' law in proteomes (a) and LEGO constructions (d) conditioned for the three largest topics.
    Proteomes are grouped at the \textit{class} taxonomic level for phylogenetic similarities, while LEGO
    constructions according to the main theme provided by the dataset annotation.
    We show here  only proteomes with size larger than 1500 and vocabulary larger than 1000 for readability.
    (b) and (e) report  the relative vocabulary,  defined as $(h(m) - \langle h \rangle(m)) / \sigma_h(m)$.
    The average and standard deviation are computed in size bins of Heaps' law and their value at a generic $m$ is obtained from a linear interpolation between them.
    (c) and (f) show the scaling of the variance with the average for the whole dataset and for realizations conditioned on a topic.
    }
    \label{fig:data_topic}
\end{figure}

\subsection{Topic-dependent variability in genomes and LEGOs}
\label{sec:topic_data}

Some datasets provide an  independent annotation of  their internal structure, offering a natural opportunity to test the latent-variable hypothesis. LEGO constructions are annotated by themes, which are closely related to the intended function of the final object and therefore to the repertoire of bricks that are likely to be used. Similarly, bacterial genomes are organized in taxonomic groups, providing a proxy for shared evolutionary history, and thus, at least partially, ecological niches and   biological functions. Although these annotations do not necessarily coincide with the latent variables of our model, they are expected to capture part of the underlying heterogeneity.

Fig. \ref{fig:data_topic} shows 
that conditioning on these annotations separates the realizations into groups with systematically different diversity levels at a given size, while preserving an approximately power-law Heaps' law within each group. 
For example, the \quotes{City LEGOs} have a larger repertoire of bricks than \quotes{Star Wars LEGOs} or  \quotes{Technic LEGOs}, but their corresponding Heaps' laws appear to follow power laws with similar exponents and different coefficients, as expected from a topic-dependent prefactor as in Eq.~\ref{eq:avh_at_topic}.   

The separation of theme-dependent vocabulary sizes can be tested  more systematically using  the \textit{relative vocabulary}, which is the distance from the ensemble average $\langle h \rangle$ in units of standard deviations $\sigma_h$, i.e., 
for each realization with $h$ and $m$  we can evaluate  $(h - \langle h \rangle(m)) / \sigma_h(m)$.
This method clearly highlights the diversity   differences between topics  across all sizes (panels b and e). 

This behavior is consistent with the latent-variable framework, in which different latent states are characterized by different Heaps' laws.  At the same time, growth models are not necessarily ruled out. A growth model in which the innovation probability depends explicitly on the same hidden variables, as in  an example previously discussed,  would produce the same behavior. 

Likewise, conditioning on the available annotations does not eliminate the observed quadratic fluctuation scaling (Fig.~\ref{fig:data_topic}e and f). This  may simply reflect the fact that the available annotations capture only part of the latent structure. In genomes, for example, phylogeny is intrinsically hierarchical and species belonging to the same annotated clade may still differ substantially in ecological adaptations and functional repertoires. Similarly, LEGO themes represent only a coarse description of the intended constructions. Consequently, lower-level or unobserved latent variables are expected to contribute additional diversity fluctuations, but data scarcity does not allow the reliable estimation of fluctuation scalings for too fine-grained annotations (when present).
In the case of genome evolution, the results are also compatible with a scenario in which a larger protein repertoire at a given point in evolutionary history unlocked the possibility of acquiring new functions (and the corresponding components) along subsequent evolutionary lineages, consistent with a ``diversity begets diversity" principle.

These observations therefore do not clearly  discriminate between the two frameworks. Rather, they show that known topic heterogeneity influences and, at least  partially,  explains the observed diversity as expected from model descriptions based on latent variables.
Moreover, the latent-variable framework provides in general a more parsimonious explanation because quadratic fluctuation scaling follows directly from heterogeneity through the law of total variance, without requiring a specific diversity-dependent innovation rule.

\section*{Discussion}

Empirically, two remarkably robust statistical regularities characterize diversity in component systems: a sublinear trend of the repertoire of distinct components with respect to the total number of components described by Heaps' law and the quadratic scaling of diversity fluctuations, a Taylor's law for diversity, revealing the non-self-averaging nature of diversity across these systems. Here, we have asked how these remarkable features emerge.
Our central result is that while they can be signatures of specific  underlying processes, the same statistical laws can emerge from two fundamentally different mechanisms.

The first is an intrinsic feedback mechanism in dynamical innovation processes, in which the probability of introducing a new component increases with existing diversity. 
This diversity-generates-diversity mechanism embodies the intuitive idea that more diverse systems are more likely to generate further diversity, a concept closely related to the evocative idea of the expansion of the adjacent possible in innovation theories~\cite{kauffman1995home,kauffman2000investigations,tria2014dynamics}. 
Notably, our analytical results show that this dependence is far from generic: among the large family of innovation models, only a narrow class reproduces the observed scaling laws. In particular, the innovation probability has to asymptotically satisfy the linear dependence $p_n \sim h/m$ which a priori is not universal, unless one finds compelling reasons why this scaling could be an attractor of more general dynamics.

In the second route,
 the observed diversity statistics emerge from structural hidden/latent heterogeneity across realizations, rather than arising from explicit innovation dynamics.  When systems are composed of mixtures of latent states, variability between these hidden states alone can generate the same macroscopic  signatures.
These latent variables may  represent structural or semantic features that affect component frequencies, such as  topics in texts or functional constraints and  environmental factors in  genomes. 
The idea that empirical  macroscopic statistical laws may emerge from hidden latent factors has previously been proposed to explain the ubiquitous presence of Zipf's law~\cite{aitchison2016zipf,schwab2014zipf}. Here we extend this perspective from component frequencies  to diversity statistics by identifying the general conditions under which latent heterogeneity reproduces both Heaps' law and its quadratic fluctuation scaling. 
Together, these results suggest a  unified statistical picture in which latent structure accounts for component frequencies, average diversity, and diversity fluctuations.

The two explanations are unlikely to be equally appropriate for every component system. Some systems, such as genomes, naturally admit an interpretation in terms of sequential growth and innovation. Others are less naturally described in this way. LEGO constructions are not designed brick by brick through an innovation process, and  gene expression profiles in single cells, which again share the same diversity statistics~\cite{lazzardi2023emergent},  are not assembled one transcript at a time. In these systems, latent variables associated with functions, themes, developmental programs, or environmental conditions appear to provide a more natural description of the observed diversity. 
At the same time, our empirical analysis of genomes and LEGO constructions shows that conditioning on known annotations isolates topic-dependent average Heaps' laws,  as expected from the latent-variable framework. However,  residual quadratic fluctuations remain. This may indicate that the available annotations capture only part of the latent structure, that the latent organization is itself hierarchical (as is naturally the case for species taxonomies), or that an innovation process where diversity begets diversity is also present. Consequently, these observations,  while confirming a key role of latent variables for diversity, are not sufficient to uniquely discriminate between the two mechanisms.

Modern large language models provide an interesting intermediate case. Despite generating text sequentially and without an explicit topic structure under random prompting, they exhibit the same effective innovation law predicted by the growth framework. This suggests that linear diversity-dependent innovation is a robust feature of autoregressive language generation by LLMs. However, this observation does not necessarily imply that these models implement an explicit innovation rule. An alternative interpretation is that  language generation itself relies on a superposition of latent modes acquired during training. Even under random prompting, the prompt may act as a selector over these hidden factors, giving rise to  linear diversity-dependent innovation as an effective  statistical property.

On theoretical grounds, the correspondence between dynamical  and latent-variable descriptions is reminiscent of representation theorems for exchangeable processes, such as  De Finetti's theorem~\cite{de2017theory,pitman2006combinatorial}, in which apparent statistical dependencies emerge from averaging over hidden variables. Our results reveal a similar conceptual duality in the context of diversity in component systems, although a rigorous mathematical formalization  of the link with representations of exchangeable processes is still missing.

 If Heaps' law and its fluctuation scaling do not uniquely identify the  mechanism responsible for their emergence, selecting the most likely scenario between an expanding state-space driven by innovation and a static but heterogeneous landscape of possibilities remains an open challenge in specific empirical systems. This ambiguity provides a warning against interpreting diversity scaling as evidence of a specific innovation process. Conversely, our findings suggest a unified view of diversity variability, where at least in terms of diversity and its fluctuations,  qualitatively different generative processes appear to belong to the same phenomenological  universality class. 


Identifying observables capable of discriminating   between alternative  underlying mechanisms remains an important direction for future work. Promising candidates include correlations and higher-order statistics~\cite{sireci2023environmental,di2025dynamics} or systems in which the relevant latent variables can be  independently and fully characterized.

\section{Code availability}

The code associated with this paper will be made publicly available on GitHub upon publication of the article.

\section{Acknowledgements}

This work was supported by the Italian ``Ministero dell’Università e della Ricerca'', PRIN 2022, COD. 2022PY8MHN, ``Genomic Component Systems (GeCoS)''.\\
We acknowledge the EuroHPC Joint Undertaking for awarding us access to 
MareNostrum5 GPP at BSC, Spain, COD. EHPC-BEN-2025B10-022.

\bibliographystyle{unsrt}
\bibliography{bib}

@incollection{page2010diversity,
  title={Diversity and complexity},
  author={Page, Scott},
  booktitle={Diversity and complexity},
  year={2010},
  publisher={Princeton University Press}
}

@article{altmann2016statistical,
  title={Statistical laws in linguistics},
  author={Altmann, Eduardo G and Gerlach, Martin},
  journal={Creativity and universality in language},
  pages={7--26},
  year={2016},
  publisher={Springer}
}

@article{angelini2010mean,
  title={Mean-field methods in evolutionary duplication-innovation-loss models for the genome-level repertoire of protein domains},
  author={Angelini, A and Amato, A and Bianconi, G and Bassetti, B and Lagomarsino, M Cosentino},
  journal={Physical Review E},
  volume={81},
  number={2},
  pages={021919},
  year={2010},
  publisher={APS}
}

@article{aitchison2016zipf,
  title={Zipf’s law arises naturally when there are underlying, unobserved variables},
  author={Aitchison, Laurence and Corradi, Nicola and Latham, Peter E},
  journal={PLoS computational biology},
  volume={12},
  number={12},
  pages={e1005110},
  year={2016},
  publisher={Public Library of Science San Francisco, CA USA}
}

@misc{attardiWiki,
  howpublished = "\url{http://attardi.github.io/wikiextractor/}"
}

@article{bassetti2009statistical,
  title={Statistical mechanics of the “Chinese restaurant process”: Lack of self-averaging, anomalous finite-size effects, and condensation},
  author={Bassetti, Bruno and Zarei, Mina and Cosentino Lagomarsino, Marco and Bianconi, Ginestra},
  journal={Physical Review E—Statistical, Nonlinear, and Soft Matter Physics},
  volume={80},
  number={6},
  pages={066118},
  year={2009},
  publisher={APS}
}

@book{bird2009nltk,
  title = {Natural Language Processing with Python: Analyzing Text with the Natural Language Toolkit},
  author = {Bird, Steven and Klein, Ewan and Loper, Edward},
  year = {2009},
  month = {6},
  publisher = {O'Reilly Media, Inc.},
  url = {https://www.nltk.org/book/},
  isbn = {9780596516499}
}

@article{blei2003latent,
  title={Latent dirichlet allocation},
  author={Blei, David M and Ng, Andrew Y and Jordan, Michael I},
  journal={Journal of machine Learning research},
  volume={3},
  number={Jan},
  pages={993--1022},
  year={2003}
}

@article{corominas2015understanding,
  title={Understanding scaling through history-dependent processes with collapsing sample space},
  author={Corominas-Murtra, Bernat and Hanel, Rudolf and Thurner, Stefan},
  journal={Proceedings of the National Academy of Sciences},
  volume={112},
  number={17},
  pages={5348--5353},
  year={2015},
  publisher={National Academy of Sciences}
}

@article{cosentino2009universal,
  title={Universal features in the genome-level evolution of protein domains},
  author={Cosentino Lagomarsino, Marco and Sellerio, Alessandro L and Heijning, Philip D and Bassetti, Bruno},
  journal={Genome biology},
  volume={10},
  number={1},
  pages={R12},
  year={2009},
  publisher={Springer}
}

@article{di2025dynamics,
  title={The dynamics of higher-order novelties},
  author={Di Bona, Gabriele and Bellina, Alessandro and De Marzo, Giordano and Petralia, Angelo and Iacopini, Iacopo and Latora, Vito},
  journal={Nature Communications},
  volume={16},
  number={1},
  pages={393},
  year={2025},
  publisher={Nature Publishing Group UK London}
}

@article{eisler2008fluctuation,
  title={Fluctuation scaling in complex systems: Taylor's law and beyond},
  author={Eisler, Zolt{\'a}n and Bartos, Imre and Kert{\'e}sz, J{\'a}nos},
  journal={Advances in Physics},
  volume={57},
  number={1},
  pages={89--142},
  year={2008},
  publisher={Taylor \& Francis}
}

@article{eliazar2011growth,
  title={The growth statistics of Zipfian ensembles: Beyond Heaps’ law},
  author={Eliazar, Iddo},
  journal={Physica A: Statistical Mechanics and its Applications},
  volume={390},
  number={20},
  pages={3189--3203},
  year={2011},
  publisher={Elsevier}
}

@article{erwin2015novelty,
  title={Novelty and innovation in the history of life},
  author={Erwin, Douglas H},
  journal={Current Biology},
  volume={25},
  number={19},
  pages={R930--R940},
  year={2015},
  publisher={Elsevier}
}

@incollection{fagerberg2006innovation,
    author = {Fagerberg, Jan},
    isbn = {9780199286805},
    title = { Innovation: A Guide to the Literature},
    booktitle = {The Oxford Handbook of Innovation},
    publisher = {Oxford University Press},
    year = {2006},
    month = {01},
    doi = {10.1093/oxfordhb/9780199286805.003.0001},
    url = {https://doi.org/10.1093/oxfordhb/9780199286805.003.0001},
}

@article{gerlach2013stochastic,
  title={Stochastic model for the vocabulary growth in natural languages},
  author={Gerlach, Martin and Altmann, Eduardo G},
  journal={Physical Review X},
  volume={3},
  number={2},
  pages={021006},
  year={2013},
  publisher={APS}
}

@article{gerlach2014scaling,
  title={Scaling laws and fluctuations in the statistics of word frequencies},
  author={Gerlach, Martin and Altmann, Eduardo G},
  journal={New Journal of Physics},
  volume={16},
  number={11},
  pages={113010},
  year={2014},
  publisher={IOP Publishing}
}

@article{giometto2015sample,
  title={Sample and population exponents of generalized Taylor’s law},
  author={Giometto, Andrea and Formentin, Marco and Rinaldo, Andrea and Cohen, Joel E and Maritan, Amos},
  journal={Proceedings of the National Academy of Sciences},
  volume={112},
  number={25},
  pages={7755--7760},
  year={2015},
  publisher={National Academy of Sciences}
}

@article{grilli2020macroecological,
  title={Macroecological laws describe variation and diversity in microbial communities},
  author={Grilli, Jacopo},
  journal={Nature communications},
  volume={11},
  number={1},
  pages={4743},
  year={2020},
  publisher={Nature Publishing Group UK London}
}

@book{heaps1978information,
  title={Information retrieval: Computational and theoretical aspects},
  author={Heaps, Harold Stanley},
  year={1978},
  publisher={Academic Press, Inc.}
}

@article{herdan1960type,
  title={Type-token mathematics: A textbook of mathematical linguistics},
  author={Herdan, Gustav},
  journal={(No Title)},
  year={1960}
}

@misc{hochberg2017innovation,
  title={Innovation: an emerging focus from cells to societies},
  author={Hochberg, Michael E and Marquet, Pablo A and Boyd, Robert and Wagner, Andreas},
  journal={Philosophical Transactions of the Royal Society B: Biological Sciences},
  volume={372},
  number={1735},
  pages={20160414},
  year={2017},
  publisher={The Royal Society}
}

@misc{holehouse2025generativemodelfunctiongrowth,
      title={A generative model of function growth explains hidden self-similarities across biological and social systems}, 
      author={James Holehouse and S. Redner and Vicky Chuqiao Yang and P. L. Krapivsky and Jose Ignacio Arroyo and Geoffrey B West and Chris Kempes and Hyejin Youn},
      year={2025},
      eprint={2509.14468},
      archivePrefix={arXiv},
      primaryClass={physics.soc-ph},
      url={https://arxiv.org/abs/2509.14468}, 
}

@article{hunter1998value,
  title={The value of microbial diversity},
  author={Hunter-Cevera, Jennie C},
  journal={Current Opinion in Microbiology},
  volume={1},
  number={3},
  pages={278--285},
  year={1998},
  publisher={Elsevier}
}

@article{iacopini2018network,
  title={Network dynamics of innovation processes},
  author={Iacopini, Iacopo and Milojevi{\'c}, Sta{\v{s}}a and Latora, Vito},
  journal={Physical review letters},
  volume={120},
  number={4},
  pages={048301},
  year={2018},
  publisher={APS}
}

@article{karev2002birth,
  title={Birth and death of protein domains: a simple model of evolution explains power law behavior},
  author={Karev, Georgy P and Wolf, Yuri I and Rzhetsky, Andrey Y and Berezovskaya, Faina S and Koonin, Eugene V},
  journal={BMC evolutionary biology},
  volume={2},
  number={1},
  pages={18},
  year={2002},
  publisher={BioMed Central}
}

@book{kauffman2000investigations,
  title={Investigations},
  author={Kauffman, Stuart A},
  year={2000},
  publisher={Oxford University Press}
}

@book{kingsley1935psycho,
  title={The psycho-biology of language: an introduction to dynamic philology},
  author={George Kingsley Zipf},
  year={1935},
  publisher={Houghton Mifflin}
}

@article{koch2007software,
  title={Software evolution in open source projects—a large-scale investigation},
  author={Koch, Stefan},
  journal={Journal of Software Maintenance and Evolution: Research and Practice},
  volume={19},
  number={6},
  pages={361--382},
  year={2007},
  publisher={Wiley Online Library}
}

@misc{gutenberg,
  howpublished = "\url{http://www.gutenberg.org}"
}

@inproceedings{lahiri2014complexity,
  title={Complexity of word collocation networks: A preliminary structural analysis},
  author={Lahiri, Shibamouli},
  booktitle={Proceedings of the student research workshop at the 14th conference of the European chapter of the association for computational linguistics},
  pages={96--105},
  year={2014}
}

@ARTICLE{lagomarsino2009universal,
  author = {Lagomarsino, Marco Cosentino and Sellerio, Alessandro L and Heijning,
	Philip D and Bassetti, Bruno},
  title = {Universal features in the genome-level evolution of protein domains},
  journal = {Genome biology},
  year = {2009},
  volume = {10},
  pages = {R12},
  number = {1},
  publisher = {BioMed Central}
}

@article{lazzardi2023emergent,
  title={Emergent statistical laws in single-cell transcriptomic data},
  author={Lazzardi, Silvia and Valle, Filippo and Mazzolini, Andrea and Scialdone, Antonio and Caselle, Michele and Osella, Matteo},
  journal={Physical Review E},
  volume={107},
  number={4},
  pages={044403},
  year={2023},
  publisher={APS}
}

@article{locey2016scaling,
  title={Scaling laws predict global microbial diversity},
  author={Locey, Kenneth J and Lennon, Jay T},
  journal={Proceedings of the National Academy of Sciences},
  volume={113},
  number={21},
  pages={5970--5975},
  year={2016},
  publisher={National Academy of Sciences}
}

@article{lu2010zipf,
  title={Zipf's law leads to Heaps' law: Analyzing their relation in finite-size systems},
  author={L{\"u}, Linyuan and Zhang, Zi-Ke and Zhou, Tao},
  journal={PloS one},
  volume={5},
  number={12},
  pages={e14139},
  year={2010},
  publisher={Public Library of Science San Francisco, USA}
}

@article{mazzolini2018heaps,
  title={Heaps' law, statistics of shared components, and temporal patterns from a sample-space-reducing process},
  author={Mazzolini, Andrea and Colliva, Alberto and Caselle, Michele and Osella, Matteo},
  journal={Physical Review E},
  volume={98},
  number={5},
  pages={052139},
  year={2018},
  publisher={APS}
}

@article{mazzolini2018zipf,
  title={Zipf and Heaps laws from dependency structures in component systems},
  author={Mazzolini, Andrea and Grilli, Jacopo and De Lazzari, Eleonora and Osella, Matteo and Lagomarsino, Marco Cosentino and Gherardi, Marco},
  journal={Physical review E},
  volume={98},
  number={1},
  pages={012315},
  year={2018},
  publisher={APS}
}

@article{mazzolini2026componentsystems,
  title = {Component systems: Do null models explain everything?},
  volume = {3},
  ISSN = {2837-8830},
  url = {http://dx.doi.org/10.1371/journal.pcsy.0000114},
  DOI = {10.1371/journal.pcsy.0000114},
  number = {6},
  journal = {PLOS Complex Systems},
  publisher = {Public Library of Science (PLoS)},
  author = {Mazzolini,  Andrea and Corigliano,  Mattia and Droghetti,  Rossana and Osella,  Matteo and Cosentino Lagomarsino,  Marco},
  editor = {Poletto,  Chiara},
  year = {2026},
  month = June,
  pages = {e0000114}
}

@article{van2003scaling,
  title={Scaling laws in the functional content of genomes},
  author={van Nimwegen, Erik},
  journal={Trends in genetics},
  volume={19},
  number={9},
  pages={479--484},
  year={2003},
  publisher={Elsevier}
}

@ARTICLE{Orengo2005,
  author = {Orengo, Christine A. and Thornton, Janet M.},
  title = {Protein families and their evolution-a structural perspective.},
  journal = {Annu Rev Biochem},
  year = {2005},
  volume = {74},
  pages = {867--900},
}

@article{pitman1997two,
  title={The two-parameter Poisson-Dirichlet distribution derived from a stable subordinator},
  author={Pitman, Jim and Yor, Marc},
  journal={The Annals of Probability},
  pages={855--900},
  year={1997},
  publisher={JSTOR}
}

@techreport{pitman1996combinatorial,
  title={Combinatorial stochastic processes},
  author={Pitman, Jim and others},
  year={1996},
  institution={Technical Report 621, Department of Statistics, University of California at Berkeley, 2002. Lecture notes for St. Flour Summer School}
}

@article{quigley1998urban,
  title={Urban diversity and economic growth},
  author={Quigley, John M},
  journal={Journal of Economic perspectives},
  volume={12},
  number={2},
  pages={127--138},
  year={1998},
  publisher={American Economic Association}
}

@article{radford2019language,
  title={Language Models are Unsupervised Multitask Learners},
  author={Radford, Alec and Wu, Jeff and Child, Rewon and Luan, David and Amodei, Dario and Sutskever, Ilya},
  year={2019}
}

@misc{rebrickable,
  howpublished = "\url{http://rebrickable.com/}"
}

@article{rosanova2017modelling,
  title={Modelling the evolution of transcription factor binding preferences in complex eukaryotes},
  author={Rosanova, Antonio and Colliva, Alberto and Osella, Matteo and Caselle, Michele},
  journal={Scientific Reports},
  volume={7},
  year={2017},
  publisher={Nature Publishing Group}
}

@article{schwab2014zipf,
  title={Zipf’s law and criticality in multivariate data without fine-tuning},
  author={Schwab, David J and Nemenman, Ilya and Mehta, Pankaj},
  journal={Physical review letters},
  volume={113},
  number={6},
  pages={068102},
  year={2014}
}

@article{simon1955class,
  title={On a class of skew distribution functions},
  author={Simon, Herbert A},
  journal={Biometrika},
  volume={42},
  number={3/4},
  pages={425--440},
  year={1955},
  publisher={JSTOR}
}

@article{taylor1961aggregation,
  title={Aggregation, Variance and the Mean},
  author={Taylor, LR},
  journal={Nature},
  volume={189},
  number={4766},
  pages={732-735},
  year={1961},
}

@article{tria2014dynamics,
  title={The dynamics of correlated novelties},
  author={Tria, Francesca and Loreto, Vittorio and Servedio, Vito Domenico Pietro and Strogatz, Steven H},
  journal={Scientific reports},
  volume={4},
  number={1},
  pages={5890},
  year={2014},
  publisher={Nature Publishing Group UK London}
}

@article{tria2018zipf,
  title={Zipf’s, Heaps’ and Taylor’s Laws are Determined by the Expansion into the Adjacent Possible},
  author={Tria, Francesca and Loreto, Vittorio and Servedio, Vito DP},
  journal={Entropy},
  volume={20},
  number={10},
  pages={752},
  year={2018},
  publisher={MDPI}
}

@article{tria2020taylor,
  title={Taylor’s law in innovation processes},
  author={Tria, Francesca and Crimaldi, Irene and Aletti, Giacomo and Servedio, Vito DP},
  journal={Entropy},
  volume={22},
  number={5},
  pages={573},
  year={2020},
  publisher={MDPI}
}

@article{van2005formal,
  title={A formal derivation of Heaps' Law},
  author={van Leijenhorst, Dick C and Van der Weide, Th P},
  journal={Information Sciences},
  volume={170},
  number={2-4},
  pages={263--272},
  year={2005},
  publisher={Elsevier}
}

@article{whittaker1972evolution,
  title={Evolution and measurement of species diversity},
  author={Whittaker, Robert Harding},
  journal={Taxon},
  volume={21},
  number={2-3},
  pages={213--251},
  year={1972},
  publisher={Wiley Online Library}
}

@article{yang2026scaling,
  title={Scaling laws for function diversity and specialization across socioeconomic and biological complex systems},
  author={Yang, Vicky Chuqiao and Holehouse, James and Youn, Hyejin and Arroyo, Jos{\'e} Ignacio and Redner, Sidney and West, Geoffrey B and Kempes, Christopher P},
  journal={Proceedings of the National Academy of Sciences},
  volume={123},
  number={7},
  pages={e2509729123},
  year={2026},
  publisher={National Academy of Sciences}
}

@article{yule1925ii,
  title={A mathematical theory of evolution, based on the conclusions of Dr. JC Willis, FR S},
  author={Yule, George Udny},
  journal={Philosophical transactions of the Royal Society of London. Series B, containing papers of a biological character},
  volume={213},
  number={402-410},
  pages={21--87},
  year={1925},
  publisher={The Royal Society London}
}

@misc{touvron2023llama2openfoundation,
      title={Llama 2: Open Foundation and Fine-Tuned Chat Models}, 
      author={Hugo Touvron and Louis Martin and Kevin Stone and Peter Albert and Amjad Almahairi and Yasmine Babaei and Nikolay Bashlykov and Soumya Batra and Prajjwal Bhargava and Shruti Bhosale and Dan Bikel and Lukas Blecher and Cristian Canton Ferrer and Moya Chen and Guillem Cucurull and David Esiobu and Jude Fernandes and Jeremy Fu and Wenyin Fu and Brian Fuller and Cynthia Gao and Vedanuj Goswami and Naman Goyal and Anthony Hartshorn and Saghar Hosseini and Rui Hou and Hakan Inan and Marcin Kardas and Viktor Kerkez and Madian Khabsa and Isabel Kloumann and Artem Korenev and Punit Singh Koura and Marie-Anne Lachaux and Thibaut Lavril and Jenya Lee and Diana Liskovich and Yinghai Lu and Yuning Mao and Xavier Martinet and Todor Mihaylov and Pushkar Mishra and Igor Molybog and Yixin Nie and Andrew Poulton and Jeremy Reizenstein and Rashi Rungta and Kalyan Saladi and Alan Schelten and Ruan Silva and Eric Michael Smith and Ranjan Subramanian and Xiaoqing Ellen Tan and Binh Tang and Ross Taylor and Adina Williams and Jian Xiang Kuan and Puxin Xu and Zheng Yan and Iliyan Zarov and Yuchen Zhang and Angela Fan and Melanie Kambadur and Sharan Narang and Aurelien Rodriguez and Robert Stojnic and Sergey Edunov and Thomas Scialom},
      year={2023},
      eprint={2307.09288},
      archivePrefix={arXiv},
      primaryClass={cs.CL},
      url={https://arxiv.org/abs/2307.09288}, 
}

@article{llama3modelcard,
  title={Llama 3 Model Card},
  author={AI@Meta},
  year={2024},
  url = {https://github.com/meta-llama/llama3/blob/main/MODEL_CARD.md}
}

@misc{wikiDumps,
  howpublished = "\url{https://dumps.wikimedia.org/enwiki/}"
}

@inproceedings{wolf2020transformers,
  title = {Transformers: State-of-the-Art Natural Language Processing},
  author = {Wolf, Thomas and Debut, Lysandre and Sanh, Victor and Chaumond, Julien and Delangue, Clement and Moi, Anthony and Cistac, Perric and Ma, Clara and Jernite, Yacine and Plu, Julien and Xu, Canwen and Le Scao, Teven and Gugger, Sylvain and Drame, Mariama and Lhoest, Quentin and Rush, Alexander M.},
  booktitle = {Proceedings of the 2020 Conference on Empirical Methods in Natural Language Processing: System Demonstrations},
  month = {10},
  year = {2020},
  publisher = {Association for Computational Linguistics},
  address = {Online},
  pages = {38--45},
  url = {https://www.aclweb.org/anthology/2020.emnlp-demos.6}
}

@article{zanette2005dynamics,
  title={Dynamics of text generation with realistic Zipf's distribution},
  author={Zanette, Dami{\'a}n and Montemurro, Marcelo},
  journal={Journal of quantitative Linguistics},
  volume={12},
  number={1},
  pages={29--40},
  year={2005},
  publisher={Taylor \& Francis}
}

@article{paysanlafosse2024pfam,
    author = {Paysan-Lafosse, Typhaine and Andreeva, Antonina and Blum, Matthias and Chuguransky, Sara Rocio and Grego, Tiago and Pinto, Beatriz Lazaro and Salazar, Gustavo A and Bileschi, Maxwell L and Llinares-López, Felipe and Meng-Papaxanthos, Laetitia and Colwell, Lucy J and Grishin, Nick V and Schaeffer, R Dustin and Clementel, Damiano and Tosatto, Silvio C E and Sonnhammer, Erik and Wood, Valerie and Bateman, Alex},
    title = {The Pfam protein families database: embracing AI/ML},
    journal = {Nucleic Acids Research},
    volume = {53},
    number = {D1},
    pages = {D523-D534},
    year = {2024},
    month = {11},
    issn = {1362-4962},
    doi = {10.1093/nar/gkae997},
    url = {https://doi.org/10.1093/nar/gkae997},
    eprint = {https://academic.oup.com/nar/article-pdf/53/D1/D523/60667756/gkae997.pdf},
}

@article{ahmad2025uniprotkb,
    author = {Ahmad, Shadab and Jose da Costa Gonzales, Leonardo and Bowler-Barnett, Emily H and Rice, Daniel L and Kim, Minjoon and Wijerathne, Supun and Luciani, Aurélien and Kandasaamy, Swaathi and Luo, Jie and Watkins, Xavier and Turner, Edd and Martin, Maria J and the UniProt Consortium },
    title = {The UniProt website API: facilitating programmatic access to protein knowledge},
    journal = {Nucleic Acids Research},
    volume = {53},
    number = {W1},
    pages = {W547-W553},
    year = {2025},
    month = {05},
    issn = {1362-4962},
    doi = {10.1093/nar/gkaf394},
    url = {https://doi.org/10.1093/nar/gkaf394},
    eprint = {https://academic.oup.com/nar/article-pdf/53/W1/W547/63079860/gkaf394.pdf},
}

@article{schoch2020ncbitaxonomy,
    author = {Schoch, Conrad L and Ciufo, Stacy and Domrachev, Mikhail and Hotton, Carol L and Kannan, Sivakumar and Khovanskaya, Rogneda and Leipe, Detlef and Mcveigh, Richard and O’Neill, Kathleen and Robbertse, Barbara and Sharma, Shobha and Soussov, Vladimir and Sullivan, John P and Sun, Lu and Turner, Seán and Karsch-Mizrachi, Ilene},
    title = {NCBI Taxonomy: a comprehensive update on curation, resources and tools},
    journal = {Database},
    volume = {2020},
    pages = {baaa062},
    year = {2020},
    month = {08},
    issn = {1758-0463},
    doi = {10.1093/database/baaa062},
    url = {https://doi.org/10.1093/database/baaa062},
    eprint = {https://academic.oup.com/database/article-pdf/doi/10.1093/database/baaa062/33570620/baaa062.pdf},
}

@book{kauffman1995home,
  title={At home in the universe: The search for laws of self-organization and complexity},
  author={Kauffman, Stuart A},
  year={1995},
  publisher={Oxford university press}
}

@book{pitman2006combinatorial,
  title={Combinatorial stochastic processes: Ecole d'et{\'e} de probabilit{\'e}s de saint-flour xxxii-2002},
  author={Pitman, Jim},
  year={2006},
  publisher={Springer Science \& Business Media}
}

@book{de2017theory,
  title={Theory of probability: A critical introductory treatment},
  author={De Finetti, Bruno},
  year={2017},
  publisher={John Wiley \& Sons}
}

@article{sireci2023environmental,
  title={Environmental fluctuations explain the universal decay of species-abundance correlations with phylogenetic distance},
  author={Sireci, Matteo and Mu{\~n}oz, Miguel A and Grilli, Jacopo},
  journal={Proceedings of the National Academy of Sciences},
  volume={120},
  number={37},
  pages={e2217144120},
  year={2023},
  publisher={National Academy of Sciences}
}

\newpage
\onecolumngrid
\newpage

\begin{center}
	\textbf{\large SUPPLEMENTARY MATERIAL}
\end{center}

\setcounter{figure}{0}
\setcounter{section}{0}
\setcounter{table}{0}
\setcounter{equation}{0}

\renewcommand{\figurename}{Supplementary Figure}
\renewcommand{\thesection}{Supplementary Section \arabic{section}}
\renewcommand{\thefigure}{S\arabic{figure}}
\renewcommand{\thesection}{S\arabic{section}}
\renewcommand{\tablename}{Supplementary Table}
\renewcommand{\thetable}{S\arabic{table}}
\renewcommand{\theequation}{S\arabic{equation}}

\section{Fast approximated simulation of an innovation process}
\label{sec:SM_approx}

The numerical evaluation of a growth model defined by Eq. 3 of the main text can be computationally intensive because, in general, we need many parallel realizations $R$, to have accurate estimates of the Taylor exponent, and many time steps $M$ to reach the asymptotic limit.
A straightforward implementation would require $R \cdot M$ operations to build a vocabulary trajectory $h(m)$.
To reduce this number we can approximately say that, for a given number of steps $\Delta m$, the innovation probability $\pn(h,m)$ is approximately constant in $m$ and $h$.
In such a case, the generation of $\Delta m$ Bernoulli variables leading to the increment $\Delta h(m)$, can be obtained by sampling a single binomial variable:
\begin{equation*}
    \Delta h(m) \sim Binom\left( \Delta m, \pn \left(\tilde{h}, \tilde{m} \right) \right) , 
    \hspace{0.5cm} \tilde{m} = m + \frac{\Delta m}{2} ,
    \hspace{0.5cm} \tilde{h} = h(m) + \frac{h(m) - h(m-\Delta m)}{2} .
\end{equation*}
Above we consider the constant value $\tilde{m}$ as the middle point in the bin $[m, m + \Delta m]$, and $\tilde{h}$ as the middle-point estimate by using the first-order approximation of the increment at the previous step. 
Since the rate of change of the first derivative is stronger in the initial part of the trajectory, $h \sim m^\nu$, we initially generate $m_{\rm buffer}$ non-approximated steps to lower the error.

\begin{figure}[h]
    \centering
    \includegraphics[width=0.95\linewidth]{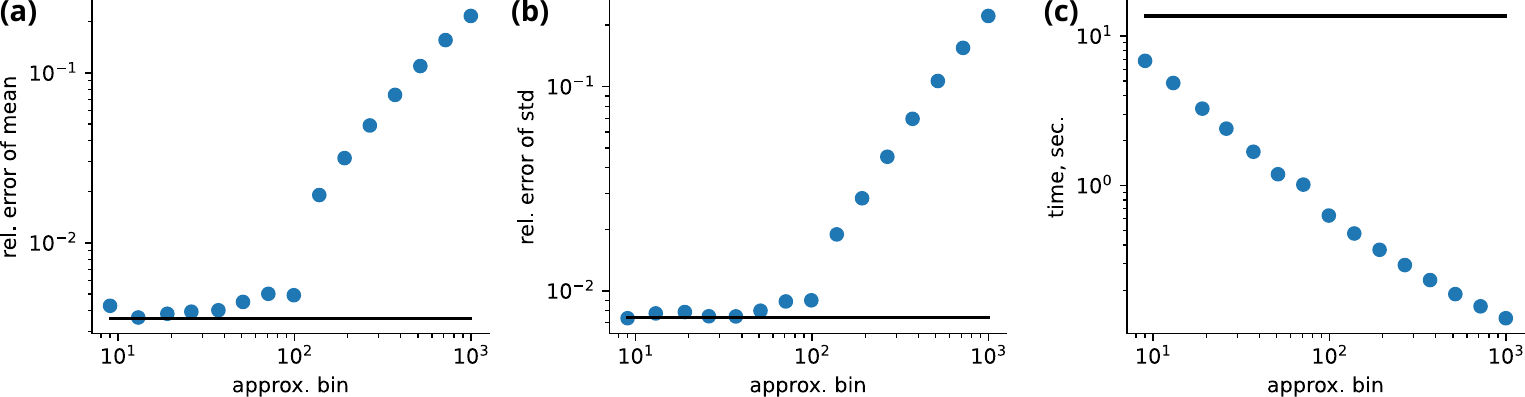}
    \caption{
    \textbf{Testing the numerical approximation for the growth model simulations.}
    Relative errors and computational time of the approximated model by varying the approximation bin size $\Delta m$.
    The relative error is defined as $\mathbb{E}_{R'} (|F - F_{true}|) / \mathbb{E}_{R'} (|F_{true}|)$, where $F$ can be the average, panel (a), or the standard deviation, panel (b), of the vocabulary $h(m)$ at the last step of the simulation $m=50000$, over an ensemble of $R=10000$ realizations.
    The average $\mathbb{E}_{R'}$ is computed for $R'=100$ repetitions of one approximated simulation (leading to F) and one non-approximated simulation (leading to $F_{true}$).
    In panels a and b, the black line shows the error that one would get with this procedure by computing $F$ with another non-approximated simulation.
    Panel c shows the average computational time across the $R'$ repetitions, and the black line is the average time on non-approximated process.
    Simulations are made with $a=0.6$, $\beta=0.8$, $\gamma=0.9$.
    }
    \label{fig:SM_approx}
\end{figure}

In Fig. \ref{fig:SM_approx}a,b we compute the errors in evaluating average and standard deviation of the vocabulary $h$ at the end of the trajectory by varying the approximation bin-size $\Delta m$.
These errors are compared with the ones of non-approximated replicates, which still lead to biases because of finite-size effects in ensembles of $R=10000$ realizations (black lines).
We can observe that, up to bin sizes of around 100 steps, the approximation performs as well as non-approximated replicates.
The panel c shows the computational time in seconds, compared with the time of the non-approximated process (black line), where, for $\Delta m \sim 100$, the approximations is around 20 times faster.
In general, for the simulations of the paper we will use  $\Delta m = 100$ and $m_{\rm buffer} = 200$.

All the scripts to generate innovation processes are contained in the repository \url{https://github.com/amazzoli/Diversity_Growth_Topics.git}.
The simulations are made in Python, which typically are vectorized across realizations.

\section{Vocabulary fluctuation scaling from a linear innovation probability}

Assuming $\pn = \alpha h / m$, with $0<\alpha<1$, we can obtain the vocabulary average growth by using Eq. 1 of the main text:
\begin{equation}
\frac{d \avh}{d m} = \alpha \frac{\avh}{m} \;\; \rightarrow \;\; \avh(m) = m^\alpha ,
\label{eq:SM_avh_lin}
\end{equation}
where we solved the equation with the initial condition $\avh(1) = 1$, i.e. at the initial time step there is only one component.
By using Eq. 2 of the main we can obtain the variance dynamics:
\begin{equation}
\frac{d \sigma_h^2}{d m} = 2 \frac{\alpha \sigma^2_h}{m}  + \alpha \frac{\avh}{m} \left( 1 - \alpha \frac{\avh}{m} \right) =  2 \frac{\alpha \sigma^2_h}{m}  + \alpha m^{\alpha-1} \left( 1 - \alpha m^{\alpha-1} \right) .
\label{eq:SM_var_h_pn_linh}
\end{equation}
This equation can be solved with the variation of the constant method, by first solving the homogeneous associated equation:
\begin{equation*}
\frac{d \hat{\sigma}_h^2}{d m} = 2 \frac{\alpha \hat{\sigma}^2_h}{m} \;\; \rightarrow \;\; \hat{\sigma}^2_h(m) = c m^{2\alpha} ,
\end{equation*}
and then writing $\sigma^2_h(m) = C(m) \hat{\sigma}_h^2(m) = C(m) m^{2\alpha}$.
By plugging this expression in Eq. \ref{eq:SM_var_h_pn_linh}, we obtain the following equation for $C(m)$:
\begin{equation*}
\frac{d C}{d m} = \alpha m^{-\alpha-1} - \alpha^2 m^{-2} \;\; \rightarrow \;\; C(m) = -\frac{\alpha^2}{m} - m^{-\alpha} + c.
\end{equation*}
By imposing the initial condition $\sigma^2_h(1) = 0$, we can fix the constant of integration $c$ and finally obtain:
\begin{equation}
\sigma_h^2(m) = (1 + \alpha^2)m^{2\alpha} - \alpha^2 m^{2\alpha - 1} - m^\alpha = (1 + \alpha^2)\avh^{2} + \alpha^2 \avh^{2 - 1/\alpha} - \avh,
\label{eq:SM_varh_lin}
\end{equation}
where the second equality expresses the variance as a function of the average vocabulary using Eq. \ref{eq:SM_avh_lin}. 
In the limit of large $m$, for $\alpha>0$, will tend to a scaling with exponent $2$ showing the non-self-averaging property.

\begin{figure}
    \centering
    \includegraphics[width=\linewidth]{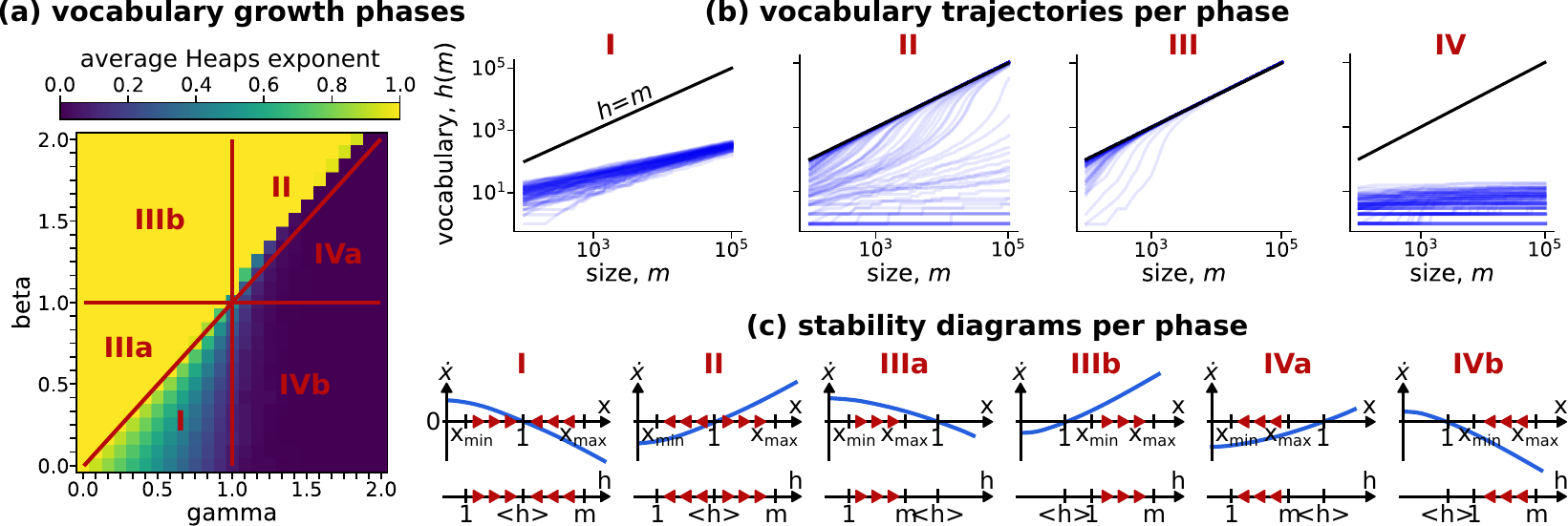}
    \caption{
    \textbf{Phases of the average Heaps' growth for different parameters of the homogeneous model.}
    The growth model is simulated with Eq. 3 varying $\beta$ and $\gamma$.
    (a) Numerical evaluation of the Heaps exponent, $\nu' = (\log \avh(m + d m) - \log\avh(m)) / (\log(m + d m) - \log m)$ at fixed $\alpha=0.5$, computed in a bin of $d m = 1000$ time steps after $10^5$ steps of $20000$ trajectories.
    The red lines separate the different phases according to the mean field analysis.
    (b) Examples of some trajectories $h(m)$ in double logarithmic scale in the four main phases: (I) $\beta=0.6$, $\gamma=0.8$, (II) $\beta=1.5$, $\gamma=1.2$, (III) $\beta=1.2$, $\gamma=0.8$ , and (IV) $\beta=0.8$, $\gamma=1.2$.
    The black line is the upper boundary $h=m$.
    (c) Sketches of the stability diagram describing qualitatively the behavior of Eq. 8.
    The special value $x=1$ is the fixed point, corresponding to $\bar{h}' \propto m^\nu$ in the $\bar{h}$ coordinates. $x_{min} = \bar{h}'^{-1}$ corresponds to the lower boundary $\bar{h}=1$ and $x_{max} = m \bar{h}'^{-1}$ to the upper boundary $\bar{h}=m$ of the dynamics.
    In the phases I and II we have $x_{min} < 1 < x_{max}$, i.e. the average $\bar{h}'$ is located within the boundaries.
    In the phase I the fixed point is stable, leading to the sub-linear growth, in the phase II is unstable, pushing the dynamics towards the boundaries.
    For all the other cases the fixed point is outside the boundary, and, depending on $\beta > 1$ or $\beta < 1$,  they will have trajectories that collapse against one of the two boundaries.
    }
    \label{fig:SM_heaps_phases}
\end{figure}

\begin{figure}
    \centering
    \includegraphics[width=0.35\linewidth]{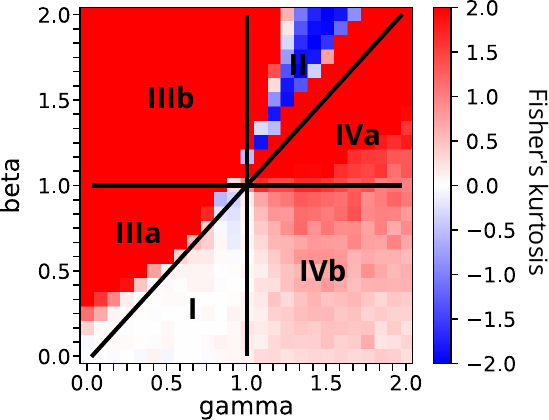}
    \caption{
    \textbf{Kurtosis of the vocabulary statistics for different parameters of the homogeneous model.}
    Numerical evaluation of the kurtosis (Fisher definition) for different parameters of the stochastic model with $\pn = \alpha h^\beta / m^\gamma$.
    Values smaller than 0 show a more peaked distribution with respect to the Gaussian, while negative values indicate a broader distribution that is possibly bimodal.
    }
    \label{fig:SM_kurtosis}
\end{figure}

\section{Non-self-averaging Heaps' law from a homogeneous innovation probability with power-law dependencies}

We study the general model with $\pn = \alpha h^\beta / m^\gamma$ in the range of parameters $\beta, \gamma < 1$ and $\gamma > \beta$.
This is the region I of Fig. \ref{fig:SM_heaps_phases} which shows a proper sub-linear power-law growth of the average.
To carry out the analytical calculations we assume that the innovation probability varies as a function of $h$ approximately as a linear function within a range $\avh \pm \sigma_h$, i.e. within the range in which the probability distribution over $h$, $P(h)$ is concentrated.
In the parameter region I, this distribution is not so distant from a Gaussian, as shown by a kurtosis close to $0$ in Fig. \ref{fig:SM_kurtosis}.
This approximation is valid if the second order term of the $\pn$ expansion is negligible with respect to the first order one in a range $\sigma_h$ around the average, i.e. $\sigma_h \partial_h \pn(\avh) \gg \sigma_h^2 \partial^2_h \pn(\avh) $.
We do not know how the variance scales with $\avh$, but one can check that this inequality consistently holds for $\avh \gg 1$ with the result found at the end of the calculation.

The first quantity we need is the average value of the innovation probability:
\begin{equation*}
    \langle \pn(h,m) \rangle = \int dh \; P(h|m) \pn(h,m) \approx \int dh \; P(h|m) \left( \pn(\avh,m) + (h - \avh) \partial_h \pn(\avh,m) \right) = \pn(\avh,m),
\end{equation*}
where $\pn$ evaluated at the average $\avh$ filters out the integral that will normalize, and the second term is equal to zero by definition of average value.
Notice that this expression corresponds to the mean field approximation used for the analysis of the average behavior, implying that can be obtained from the hypothesis of linear $\pn$.
As a consequence, we get the same average growth by solving the equation starting from $\avh(1) = 1$ obtained in Eq. 8:
\begin{equation}
    \avh = \left(\frac{\alpha}{\nu}\right)^{1/(1-\beta)} \; (m^\nu - 1) + 1 \approx \left(\frac{\alpha}{\nu}\right)^{1/(1-\beta)} \; m^\nu = k \; m^\nu,
    \label{eq:SM_avh_gen}
\end{equation}
where $\nu = (1-\gamma)/(1-\beta)$ and $k$ is defined using the last equality.

The same Taylor expansion for $\pn$ can be used for the correlation-like term of the variance equation, Eq. 2 (below we do not write the dependence on $m$ for the sake of a more compact notation):
\begin{equation*}
\begin{split}
    \langle \pn(h) h \rangle - \langle \pn(h) \rangle \avh & \approx \langle \pn(h) (h - \avh) \rangle = \int dh \; P(h) \left[ \pn(\avh) + (h - \avh) \partial_h \pn(\avh) \right] (h - \avh) = 
    \\
    & \pn(\avh) \int dh \; P(h) (h - \avh) +  \partial_h \pn(\avh) \int dh \; P(h) (h - \avh)^2 = 
    \\
    & \alpha \beta \frac{\avh^{\beta-1}}{m^\gamma} \sigma_h^2 .
\end{split}
\end{equation*}
We can finally plug the expressions found above in Eq. 2:
\begin{equation}
\frac{d \sigma_h^2}{d m} = 2 \nu \beta \frac{\sigma^2_h}{m} + \alpha \frac{\avh^\beta}{m^\gamma} \left( 1 - \alpha \frac{\avh^\beta}{m^\gamma} \right) =  2 \nu \beta \frac{\sigma^2_h}{m} + \alpha k^\beta m^{\nu \beta - \gamma} - \left( \alpha k^\beta m^{\nu \beta - \gamma} \right)^2
\label{eq:SM_var_h_pn_gen}
\end{equation}
which uses the definition of $k$ of Eq. \ref{eq:SM_avh_gen}.
Since we are interested in the scaling behavior in the asymptotic limit, to further simplify the expression we can observe that the second term will grow faster than the third one in the parameter region I, allowing us to neglect it:
\begin{equation*}
\frac{d \sigma_h^2}{d m} \approx 2 \nu \beta \frac{\sigma^2_h}{m} + \alpha k^\beta m^{\nu \beta - \gamma} .
\end{equation*}
The next step is to change the variables from $m$ to $\avh = k m^\nu$, leading to:
\begin{equation*}
\frac{d \sigma_h^2}{d \avh} \approx 2 \beta \frac{\sigma^2_h}{\avh} + 1 .
\end{equation*}
We can solve this equation with the variation of the constant and we fix the integration constant with $\sigma_h^2(1) = 0$, finding:
\begin{equation}
\begin{aligned}
&\sigma_h^2 \approx \frac{1}{1 - 2\beta} \left( \avh - \avh^{2 \beta} \right)  &\text{ for } \beta \neq 1/2\\
&\sigma_h^2 \approx \avh \log \avh &\text{ for } \beta = 1/2
\end{aligned}.
\label{eq:SM_var_h_sol_pn_gen}
\end{equation}
It is useful to obtain an analytical expression for the effective exponent $\rho$, such that $\sigma_h^2 \propto \avh^\rho$, that we can use as a comparison for numerical evaluations of this scaling.
To this end, we consider the logarithmic transformations of the two variables, $y = \log \sigma_h^2$, $x = \log \avh$, and compute the derivative $dy/dx$:
\begin{equation}
\rho = \frac{d y}{d x} = \frac{\avh - 2\beta \avh^{2 \beta}}{\avh - \avh^{2 \beta}} .
\label{eq:SM_eff_taylor_exp}
\end{equation}

\begin{figure}
    \centering
    \includegraphics[width=\linewidth]{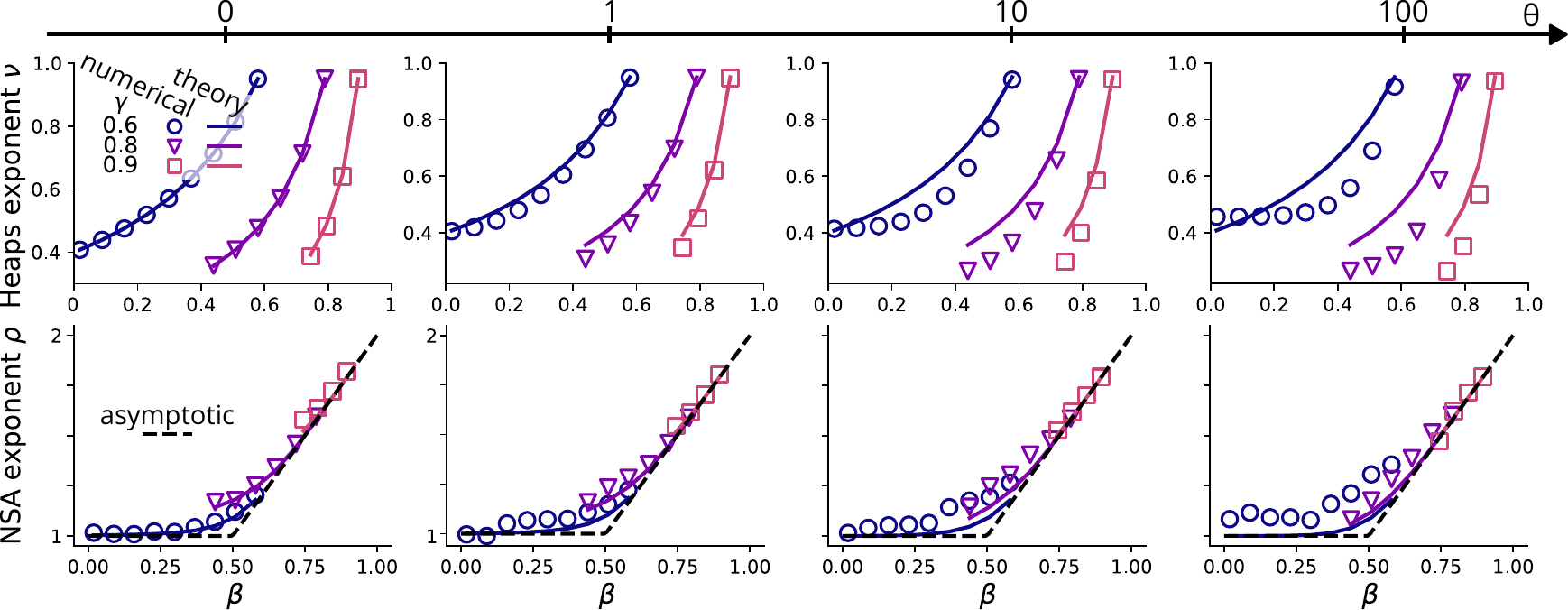}
    \caption{
    \textbf{Average Heaps and fluctuation scaling exponents for a growth model with an innovation rate with generic dependence on size and vocabulary.}
    The innovation model is defined as $\pn = (\theta + a h^\beta) / (\theta + m^\gamma)$.
    We show the same type of plot as in Fig. 2 of the main text, for increasing values of $\theta$ from the left to the right as displayed in the arrow, with the left-most plot corresponding exactly to Fig. 2.
    Simulations are performed with the same parameters reported in Fig. 2.
    }
    \label{fig:SM_taylor_theta}
\end{figure}

\section{Diversity fluctuation scaling from topic-dependent exponents of the average conditioned vocabulary growth}
\label{sec:SM_topic_expn}

Here we consider a topic-dependent vocabulary statistics $P(h) = P(h|t)P(t)$ and the following conditional average:
\begin{equation}
    \mathbb{E}_{h|t}[h] = k  \; m^{\nu(t)} .
    \label{eq:SM_av_given_topic_exp}
\end{equation}
Therefore we move the source of variability from the prefactor to the exponents of the average growth.
We analyze a case where $\nu(t)$ is a Gaussian centered in $\nu_0$ with a \quotes{small} variance $\sigma^2_\nu$ (its precise order of magnitude will be clear later).
To derive its variance with respect to the topics, we can notice that the moment generating function of a Gaussian reads:
\begin{equation*}
    \mathbb{M}_{\nu}(s) = \mathbb{E}_{\nu}[e^{s \nu}] = \exp \left[ \nu_0 s + \frac{1}{2} \sigma_\nu^2 s^2 \right] ,
\end{equation*}
which allows us to compute the global average growth:
\begin{equation}
    \avh \coloneqq \mathbb{E}_{t}[k  \; m^{\nu(t)}] = k \mathbb{E}_{t}[e^{\log m \nu(t)}] = k \mathbb{M}_{\nu}(\log m) = k m^{\nu_0} \exp \left[ \frac{1}{2} \sigma_\nu^2 (\log m)^2 \right] .
    \label{eq:SM_avh_topic_exp}
\end{equation}
By computing the second moment in a similar fashion, we can obtain:
\begin{equation*}
    \text{Var}_{t}[k  \; m^{\nu(t)}] = k^2 m^{2 \nu_0} \exp \left[  \sigma_\nu^2 (\log m)^2 \right] \left( \exp \left[  \sigma_\nu^2 (\log m)^2 \right] - 1 \right) = \avh^2 \left( \exp \left[  \sigma_\nu^2 (\log m)^2 \right] - 1 \right),
\end{equation*}
which is the second quadratic term appearing in the law of total variance Eq. 13 for the case with a variable prefactor $k(t)$.
Here the dependency is more complex because of the additional term depending on $m$, but we can reasonably assume that the exponent variance is small, such that $\sigma_\nu \log m \ll1$, leading, at the first order, to:
\begin{equation}
    \text{Var}_{t}[k  \; m^{\nu(t)}] \approx \left( \avh \sigma_\nu \log m \right)^2 \approx \left( \frac{\sigma_\nu}{\nu_0}  \avh \log \left(\avh / k\right) \right)^2 ,
    \label{eq:SM_varh_topic_exp}
\end{equation}
where the second approximate equality comes from Eq. \ref{eq:SM_avh_topic_exp}: $m \approx (\avh/k)^{1/\nu_0}$.
Therefore the scaling deviates from the NSA trend by a logarithmic correction.
The multiplying factor is the coefficient of variation of $\nu(t)$, that controls the crossover to the NSA scaling.

Regarding the average innovation probability at given topic, we can compute the counterpart of Eq. 14 for a variable $\nu(t)$, which reads:
\begin{equation}
    \mathbb{E}_{h|t}[\pn] = \nu(t) k m^{\nu(t) - 1} = \frac{\log \mathbb{E}_{h|t}[h] - \log k}{\log m} \frac{\mathbb{E}_{h|t}[h]}{m} ,
    \label{eq:SM_pnew_topic_exp}
\end{equation}
where the second equation is obtained by using Eq. \ref{eq:SM_av_given_topic_exp}.
Again, it behaves approximately as Eq. 14, showing the Diversity Generates Diversity mechanism with logarithmic corrections.

\section{Diversity fluctuations in a heterogeneous innovation process}

\subsection{Case of $\beta = 0$}

We first focus on the model without a dependency on the vocabulary, but with realizations having variable $\alpha_j$:
\begin{equation*}
    \pn = \alpha_j m^{-\gamma} .
\end{equation*}
The average vocabulary growth is obtained from Eq. 1 at fixed $\alpha$ and then averaged over the $\alpha$:
\begin{equation}
    \avh \coloneqq \mathbb{E}_\alpha[\mathbb{E}_h[h|\alpha]]  = \frac{\mathbb{E}_\alpha[\alpha]}{1-\gamma} (m^{1-\gamma} - 1) + 1 \approx \frac{\mathbb{E}_\alpha[\alpha]}{1-\gamma} m^{1-\gamma}, 
    \label{eq:SM_avh_hetero}
\end{equation}
where we use the notation with angular brackets only for the overall average across \quotes{topics} and the vocabulary.

\begin{figure}
    \centering
    \includegraphics[width=0.45\linewidth]{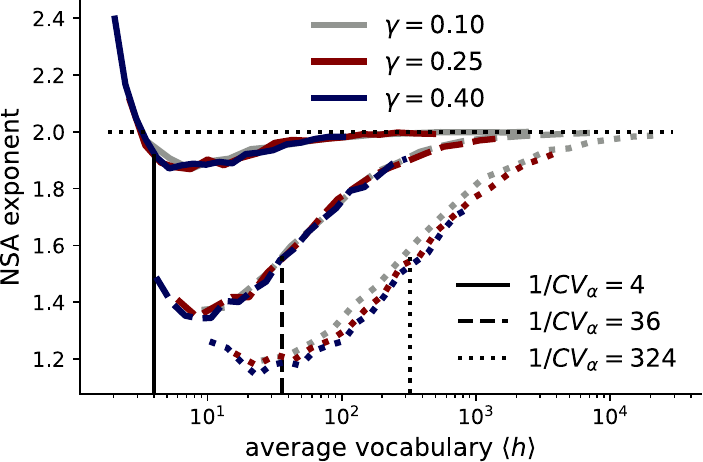}
    \caption{
    \textbf{NSA exponent convergence to 2 in a growth model with heterogeneity}
    Numerical evaluation of the vocabulary variance-average exponent, $\rho$, as a function of the average $\avh$, in a model with $\pn = \alpha_i m^{-\gamma}$.
    The parameters $\alpha_i$ are sampled from a beta distribution by fixing its standard deviation to $0.025$ and choosing three values for the average, $(0.05, 0.15, 0.45)$, that lead to the three inverse coefficients of variation in the legend.
    The black vertical lines correspond to the inverse $\text{CV}_\alpha$ used for the lines having the same style.
    For $\avh$ much larger than that value we expect the lines to converge to $2$, as shown by the figure.
    For each choice of $\text{CV}_\alpha$, we simulate trajectories for different $\gamma$ as reported in the legend.
    We simulate ensembles of $R=50000$ trajectories for $m=200000$ steps and using the numerical approximation discussed in Sec. \ref{sec:SM_approx} with $\Delta m = 50$.
    To reduce noise, trajectories have been smoothed by taking the average in a bin of $0.4$ in the log-transformed variable.
    }
    \label{fig:SM_var_mean_heter}
\end{figure}

To write the equation for the variance we have to manipulate a bit the term $\langle h \alpha \rangle$:
\begin{equation*}
    \langle h \alpha \rangle = \frac{1}{R} \sum_j \int dh P(h | \alpha_j) \;h \; \alpha_j = \frac{1}{R} \sum_j \alpha_j \mathbb{E}_h[h|\alpha_j]
    = \frac{1}{R} \sum_j \alpha_j \frac{\alpha_j}{1-\gamma} m^{1-\gamma}
    = \frac{\mathbb{E}_\alpha[\alpha^2]}{1-\gamma} m^{1-\gamma} ,
\end{equation*}
By plugging this expression in Eq. 2, we obtain:
\begin{equation*}
    \frac{d \sigma^2_h}{d m} = 2 \left( \mathbb{E}_\alpha[\alpha^2] - \mathbb{E}_\alpha[\alpha]^2\right) \frac{m^{1-2\gamma}}{1-\gamma} + \mathbb{E}_\alpha[\alpha] m^{-\gamma} - \mathbb{E}_\alpha[\alpha]^2 m^{-2 \gamma},
\end{equation*}
which can be integrated and expressed as a function of $\avh$, Eq. \ref{eq:SM_avh_hetero}:
\begin{equation}
    \sigma^2_h \approx \frac{\text{Var}_\alpha[\alpha]}{\mathbb{E}_\alpha[\alpha]^2} \avh^2 + \avh,
    \label{eq:SM_varh_hetero}
\end{equation}
where $\text{Var}_\alpha[\alpha] = \mathbb{E}_\alpha[\alpha^2] - \mathbb{E}_\alpha[\alpha]^2$ and we neglect terms of the order of $\avh^{1 - \gamma/(1-\gamma)}$.
The numerical exponent of the Taylor's law and its convergence to $2$ is shown in Fig. \ref{fig:SM_var_mean_heter}.

\subsection{General case of $\beta < 1$}

\begin{figure}
    \centering
    \includegraphics[width=\linewidth]{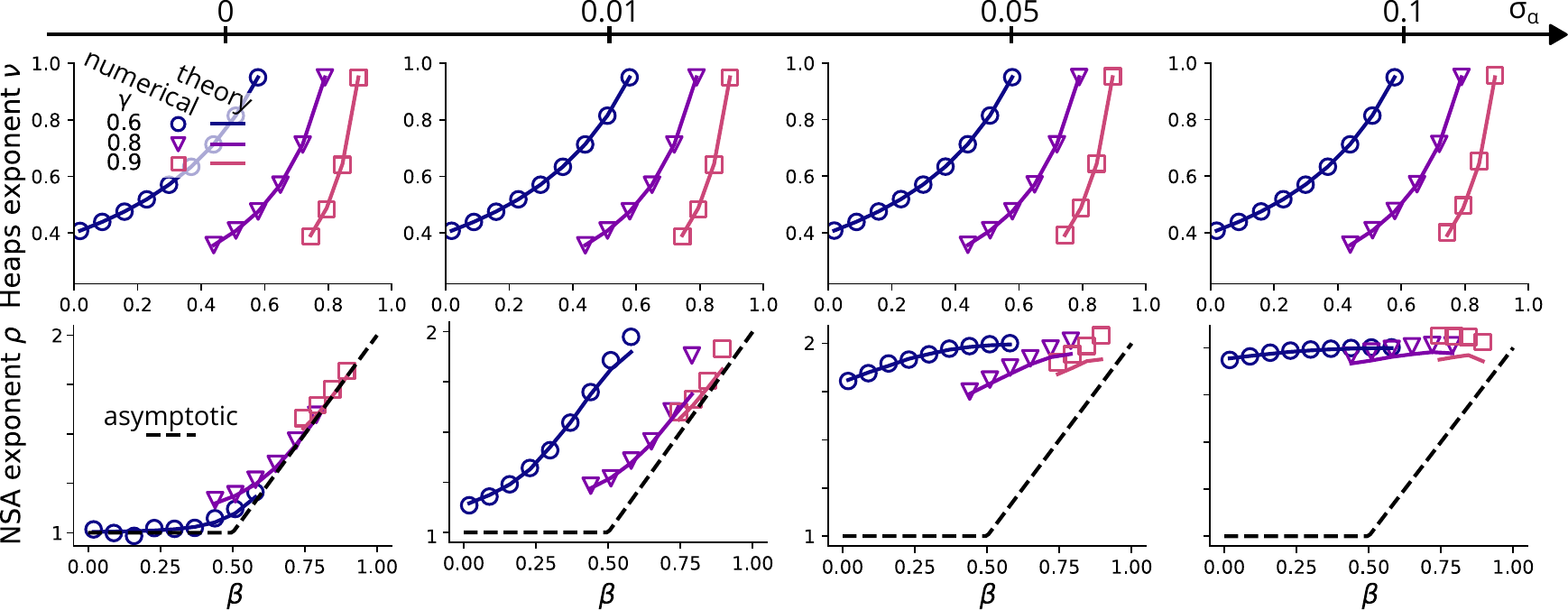}
    \caption{
    \textbf{Average Heaps and fluctuation  exponents for growth models with increasing heterogeneity.}
    As in Fig. \ref{fig:SM_taylor_theta} we show the average Heaps and the NSA exponent as a function of $\beta$ and for different $\gamma$ (colors).
    In contrast, here we consider the heterogeneous model where the factor $\alpha$ is sampled for each realization from a beta distribution whose parameters are fixed to have average $\langle \alpha \rangle = \nu = (1-\gamma)/(1-\beta)$ and increasing standard deviation as shown in the upper arrow.
    Simulation parameters are $m_{max} = 10^6$, $R=10^5$ realizations and approximation step $\Delta m=100$.
    The theoretical curves are obtained from Eqs. \ref{eq:SM_avh_hetero}, \ref{eq:varh_hetero}.
    }
    \label{fig:SM_NSA_topics}
\end{figure}

To study the general compound case of Eq. 3, we can notice that the stochastic variable $h$ can be seen as a conditioned process on the choice of $\alpha$: $P(h) = P(h|\alpha)P(\alpha)$.
The average growth can be then written as the expected value over the $\alpha$ distribution of the average growth at given $\alpha$ (given by Eq. 8),
\begin{equation}
    \avh \coloneqq \mathbb{E}_\alpha[\mathbb{E}_h[h|\alpha]] \approx \frac{\mathbb{E}_\alpha[y]}{\nu^{1/(1-\beta)}} m^\nu .
\end{equation}
where $y \coloneq \alpha^{1/(1-\beta)}$.
Therefore, we do not expect any change of the Heaps' law exponent, as shown in Fig. \ref{fig:SM_NSA_topics}c,e.

We can also use the conditioning to write the law of total variance and obtain an expression for this general case:
\begin{equation*}
    \sigma^2_h = \mathbb{E}_\alpha\left[ \sigma^2_{h|\alpha} \right] + \text{Var}_\alpha \left[ \avh_{h|\alpha} \right] .
\end{equation*}
The first contribution comes from the expected value over $\alpha$ of the variance computed at fixed $\alpha$, whose expression we know from Eq. 10.
The second contribution is given by the variance generated by the average vocabulary at different $\alpha$, i.e., the variance over $\alpha$ of the average computed in Eq. Eq. 8.
Putting all these elements together one can then write:
\begin{equation}
    \sigma^2_h \approx \frac{\text{Var}_\alpha[y]}{\mathbb{E}_\alpha[y]^2} \avh^2 + \frac{1}{2\beta-1} \left(\frac{\mathbb{E}_\alpha[y^{2\beta}]}{\mathbb{E}_\alpha[y]^{2\beta}} \avh^{2\beta} - \avh \right) ,
    \label{eq:var_generic}
\end{equation}
which recovers Eq. \ref{eq:SM_varh_hetero} for $\beta = 0$ and Eq. 10 for constant $\alpha$.
For sufficiently large $\avh$, even for a small variability in $\alpha$, the variance will eventually scale quadratically, showing a very robust mechanism to reach the empirical exponent without any need of tuning the model.
The convergence to $\rho=2$ depends on now the value of $\avh$ compares with the coefficients shown by the equation, which show quite intricate dependencies on $\beta$ and non-integer moments of $\alpha$.
Panels b, d, f in Fig. \ref{fig:SM_NSA_topics} show this crossover to the quadratic scaling for increasing choices of the standard deviation of $\alpha$.

\subsection{Case of $\beta = 1$}

In the case of $\beta=1$ the topic-conditioned average is given by Eq. \ref{eq:SM_avh_lin}, leading to:
\begin{equation*}
    \avh \coloneqq \mathbb{E}_\alpha[\mathbb{E}_h[h|\alpha]]  = \mathbb{E}_\alpha[m^\alpha] .
\end{equation*}
We now make a couple of assumptions to proceed with the computation in a simplified setting.
Specifically, we assume a Gaussian distributed $\alpha$, with average $\alpha_0$ and standard deviation $\sigma_\alpha$.
By following the same procedure of Eq. \ref{eq:SM_avh_topic_exp}, we obtain:
\begin{equation}
    \avh  = m^{\alpha_0}\; \exp \left(  \frac{\sigma_\alpha^2}{2} \log^2 m \right).
    \label{eq:SM_avh_lin_hetero}
\end{equation}
Next, we need the two terms of the law of total variance.
The variance of the conditional averages can be derived as for Eq. \ref{eq:SM_varh_topic_exp}:
\begin{equation}
    \text{Var}_{\alpha}[\mathbb{E}_h[h|\alpha]] = \text{Var}_{\alpha}[m^\alpha] \approx \left(\text{CV}[\alpha] \avh \log \avh \right)^2 ,
    \label{eq:SM_varh_lin_hetero1}
\end{equation}
where we approximated the result for a small variance $\sigma_\alpha \log m \ll 1$.

The second term is the average of the conditional variance:
\begin{equation*}
    \mathbb{E}_\alpha \left[\text{Var}_\alpha[h|\alpha] \right] = \mathbb{E}_\alpha \left[ (1+\alpha^2) m^{2\alpha} \right], 
\end{equation*}
where we used Eq. \ref{eq:SM_varh_lin}, in the limit of large sizes, to get the conditional variance.
Under the assumption of a Gaussian variable, the expectation value $\mathbb{E}_\alpha[\alpha^2 m^{2\alpha}]$ can be solved with the trick:
$$
\mathbb{E}_\alpha \left[ \alpha^2 e^{\alpha s} \right] = \frac{d^2}{ds^2} \mathbb{E}_\alpha \left[ e^{\alpha s} \right] = \left( \sigma_\alpha^2 + (\alpha_0 + s \sigma^2_\alpha)^2 \right) \mathbb{E}_\alpha \left[ e^{\alpha s} \right] ,
$$
and then substituting $s = 2 \log m$.
The final result reads:
\begin{equation}
    \mathbb{E}_\alpha \left[\text{Var}_\alpha[h|\alpha] \right] =  \left( 1 + \sigma_\alpha^2 + \left( \alpha_0 + 2 \sigma^2_\alpha \log m \right)^2 \right) \avh^2 \exp \left( 2 \sigma_\alpha^2 \log^2 m \right) \approx \left( 1 + \sigma_\alpha^2 + \alpha_0^2 \right) \avh^2
    \label{eq:SM_varh_lin_hetero2}
\end{equation}
In the limit of small topic variability, $\sigma_\alpha \log m \ll 1$, the dependencies on the size $m$ disappear, leaving only the quadratic dependency of the vocabulary.

In conclusion, by adding together the two terms of the variance, Eqs. \ref{eq:SM_varh_lin_hetero1} and \ref{eq:SM_varh_lin_hetero2}, the overall dependency on the average vocabulary is slightly more intricate than previous cases, but, for small topic variability, the quadratic scaling is recovered with a logarithmic correction.

\end{document}